\documentclass[%
 reprint,
 amsmath,amssymb,
 aps,
 prb,
 superscriptaddress,
 nobibnotes
]{revtex4-1}

\usepackage{graphicx}% Include figure files
\usepackage{dcolumn}% Align table columns on decimal point
\usepackage{bm}% bold math
\begin{document}

%\preprint{APS/123-QED}

\title{Effect of inelastic scattering
 on the nuclear magnetic relaxation rate $1/T_1T$ \\
 in iron-based superconductors}
% Force line breaks with \\

\author{Youichi \textsc{Yamakawa}}
% \email[]{yamakawa@s.phys.nagoya-u.ac.jp}
\affiliation{Department of Physics, Nagoya University and JST, TRIP, Furo-cho, Nagoya 464-8602, Japan. }
\author{Seiichiro \textsc{Onari}}
\affiliation{Department of Applied Physics, Nagoya University and JST, TRIP, Furo-cho, Nagoya 464-8603, Japan.}
\author{Hiroshi \textsc{Kontani}}
\affiliation{Department of Physics, Nagoya University and JST, TRIP, Furo-cho, Nagoya 464-8602, Japan. }

% %\collaboration{MUSO Collaboration}%\noaffiliation

% \author{Seiichiro Onari}
% %  \homepage{http://www.Second.institution.edu/~Charlie.Author}
% \affiliation{
%  Department of Applied physics, Nagoya University and JST, TRIP, Furo-cho, Nagoya 464-8602, Japan.
% }%
% % \affiliation{
% %  Third institution, the second for Charlie Author
% % }%

\date{\today}% It is always \today, today,

\begin{abstract}
We present a microscopic study of 
the nuclear magnetic relaxation rate $1/T_{1}$ 
based on the five-orbital model for iron-based superconductors.
We mainly discuss the effect of the ``inelastic'' quasi-particle damping rate 
$\gamma$ due to many-body interaction on the size of the coherence peak,
for both $s_{++}$ and $s_{\pm}$-wave superconducting states.
We focus on Ba(Fe$_{1-x}$Co$_x$)$_2$As$_2$, and systematically evaluate 
$\gamma$ in the normal state from the experimental resistivity, 
from optimally to over-doped compounds.
%add
Next, $\gamma$ in the superconducting state is calculated microscopically
based on the second order perturbation theory. 
In optimally doped compounds ($T_{\rm c} \sim 30$~K), 
it is revealed that the coherence peak on 
$1/T_{1}T$ is completely suppressed due to large $\gamma$ 
for both $s_{++}$ and $s_{\pm}$-wave states.
On the other hand, in heavily over doped compounds with $T_{\rm c} < 10$~K,
the coherence peak could appear for both pairing states,
since $\gamma$ at $T_{\rm c}$ is quickly suppressed in proportion to $T_{\rm c}^2$.
By making careful comparison between theoretical and experimental results,
we conclude that it is difficult to discriminate between 
$s_{++}$ and $s_{\pm}$-wave states from the present experimental results. 
%\begin{description}
% Secondary publications and information retrieval purposes.
% May be entered using the \verb+\pacs{#1}+ command.
% You may use the \texttt{description} environment to structure your abstract;
% use the optional argument of the \verb+\item+ command to give the category of each item. 
%\end{description}
\end{abstract}

\maketitle

%\tableofcontents
\section{Introduction}

 Since the discovery of superconductivity in LaFeAsO$_{1-x}$F$_{x}$, \cite{kamihara_iron-based_2008} 
the superconducting mechanism and paring symmetry had been discussed intensively. 
In many optimally-doped compounds, the superconducting gap is fully gapped as reported by the penetration depth measurement\cite{hashimoto_microwave_2009} and the angle-resolved photoemission spectroscopy (ARPES). \cite{ding_observation_2008, kondo_momentum_2008} 
 As for the superconducting mechanism, the $s$-wave pairing with sign change of the order parameter between the hole and electron Fermi pockets, so called $s_{\pm}$-wave, mediated by the anti-ferromagnetic fluctuation had been proposed from the early stage as a possible pairing state in the iron pnictides. \cite{kuroki_unconventional_2008, mazin_unconventional_2008,  chubukov_magnetism_2008, hirschfeld_gap_2011, cvetkovic_multiband_2009} 
It is supported by quasiparticle interference analysis by Scanning Tunneling Microscopy/Spectroscopy (STM/STS) measurement
by Hanaguri {\it et al.} \cite{Hanaguri}
 However, the small $T_{\rm c}$-suppression against nonmagnetic impurities is not consistent with the $s_{\pm}$-wave state. \cite{kawabata_superconductivity_2008, nakajima_suppression_2010, li_linear_2011, kirshenbaum_universal_2012, onari_violation_2009}

On the other hand, orbital fluctuation mediated $s$-wave superconducting state without sign reversal ($s_{++}$-wave state) had been investigated. \cite{kontani_orbital-fluctuation-mediated_2010} 
We have shown that strong ferro- and antiferro-orbital fluctuations develop
due to the combination of Coulomb and $e$-ph interactions.
\cite{onari_non-fermi-liquid_2012, saito_orbital_2010, kontani_origin_2011, onari_self-consistent_2012, kontani_orbital_2012}
Consistently, large softening of the shear modulus $C_{66}$
\cite{fernandes_effects_2010, goto_quadrupole_2011, yoshizawa_structural_2012} and 
renormalization of phonon velocity
\cite{niedziela_phonon_2011}
are observed well above the orthorhombic structure transition temperature $T_{\rm s}$.
These phenomena strongly suggest the existence of
strong ferro-orbital (charge quadrupole $O_{x^2-y^2}$) fluctuations,
considering the large strain-quadrupole coupling.
\cite{kontani_origin_2011, onari_self-consistent_2012, kontani_orbital_2012}
In addition, experimental ``resonance-like'' hump structure 
in the neutron inelastic scattering is well reproduced in terms 
of the $s_{++}$-wave state.
\cite{onari_structure_2010, onari_neutron_2011}

 The nuclear-magnetic resonance (NMR) and the nuclear-quadrupole resonance (NQR) measurements are useful for the discussion of the pairing symmetry. 
 The coherence effect of superconductivity, appearing as a Hebel-Slichter peak (coherence peak) in the nuclear spin relaxation rate ($1/T_{1}$), was one of the crucial experimental proofs of the Bardeen-Cooper-Schrieffer (BCS) theory, characterized by conventional $s$-wave Cooper pairs with an isotropic gap. \cite{hebel_nuclear_1959} 
 In the Fe based superconductors, many experimental results of $1/T_{1}$ have been published. \cite{nakai_evolution_2008, grafe_^75as_2008, kawasaki_two_2008, ning_spin_2009, yashima_strong-coupling_2009, kobayashi_studies_2009, mukuda_coherence_2010} 
It was found that the coherence peak in $1/T_{1}$ is absent
in many compounds, like electron-doped Ba(Fe$_{1-x}$Co$_{x}$)$_{2}$As$_{2}$ 
and hole-doped Ba$_{1-x}$K$_{x}$Fe$_{2}$As$_{2}$.

The size of the coherence peak had attracted great attention 
to distinguish between $s_{++}$ and $s_{\pm}$-wave states:
In a simple BCS theory, the peak size is larger
in the $s_{++}$-wave state, while it is reduced in the $s_{\pm}$-wave state.
However, the coherence peak is 
suppressed by the ``inelastic'' quasi-particle damping rate $\gamma$,
which is prominent in moderately and strongly correlated systems.
\cite{akis_damping_1991, fujimoto_many-body_1992}
For this reason, the coherence peak is not observed even in
several conventional $s$-wave superconductors with $T_{\rm c}>15$~K,
such as boron carbide YNi$_{2}$B$_{2}$C \cite{kohara_superconducting_1995} 
and A-15 compounds V$_{3}$Si. \cite{kishimoto_51v_1995} 
Thus, inelastic scattering due to many-body effect
has to be taken into account for a quantitative analysis of $1/T_1$.

Impurity effect (=``elastic'' quasi-particle damping effect)
on $T_{\rm c}$ and $1/T_1$ is also important in studying the pairing symmetry.
In the $s_{++}$-wave state, both $T_{\rm c}$ and $1/T_1$
are insensitive to impurities.
In contrast, the $s_\pm$-wave state is easily suppressed by impurities
like the $d$-wave state,
according to the study based on the five-orbital model.
\cite{onari_violation_2009}
The impurity-induced gapless state in the $s_\pm$-wave state
would give strong influence on $1/T_1$ for $T\ll T_{\rm c}$.
\cite{chubukov_magnetism_2008, bang_possible_2008, parker_extended_2008}
However, impurity-induced gapless state
is realized only when $T_{\rm c}$ is strongly suppressed to be $\sim T_{\rm c0}/3$ theoretically. \cite{senga}
Thus, gapless behavior of $1/T_1$ for $T\ll T_{\rm c}$ observed 
in some compounds cannot be explained by this scenario.

Recently, authors in Refs. 
\onlinecite{sato_studies_2010} and \onlinecite{sato_superconducting_2012}
had shown that the coherence peak in the $s_{++}$-wave state
disappears when inelastic scattering $\gamma$ at $T_{\rm c}$ is 
as large as the superconducting gap at $T=0$.
 However, quantitative estimations of $\gamma$
and its $T$ and $\omega$ dependences are still lacking.
 Interestingly, recent NMR measurement reports a small coherence peak of $1/T_{1}$ in the heavily over doped LaFeAsO$_{1-x}$F$_{x}$ with low transition temperature $T_{\rm c} \sim 5$~K.
 \cite{mukuda_coherence_2010}
 Now, microscopic study of $1/T_1$ by including
inelastic scattering effect is desired
to discuss the superconducting pairing state as $s_{\pm}$ or $s_{++}$.

 In this paper, we investigate the effect of inelastic scattering rate 
$\gamma$ on the nuclear magnetic relaxation rate $1/T_{\rm 1}$
for both $s_{++}$ and $s_{\pm}$-wave states.
For a quantitative discussion on the existence of the 
Hebel-Slichter coherence peak, we employ the two-dimensional 
five orbital model. \cite{kuroki_unconventional_2008} 
Here, $\gamma$ is the key parameter of the present study:
At $T_{\rm c}$, the value of $\gamma$
is carefully estimated from the experimental conductivity, 
\cite{onari_neutron_2011} and its temperature and the frequency 
dependences below $T_{\rm c}$ is obtained by the 
second order perturbation theory. 
 All results shown below are obtained with band filling $n = 6.1$.

 The contents of this paper are as follows; 
 In Sec. \ref{sec:formulation}, we explain how to calculate the
 nuclear magnetic relaxation rate $1/T_{1}T$ and 
quasi-particle damping $\gamma$. 
The obtained numerical results are explained in Sec. \ref{sec:results}.
 Finally, we make comparison between theoretical and experimental
results in Sec. \ref{sec:summary}, and discuss the possible pairing symmetry.

\section{Formulation}\label{sec:formulation}

\subsection{Green function}
\begin{figure}[tb]
 	\includegraphics[width=6cm]{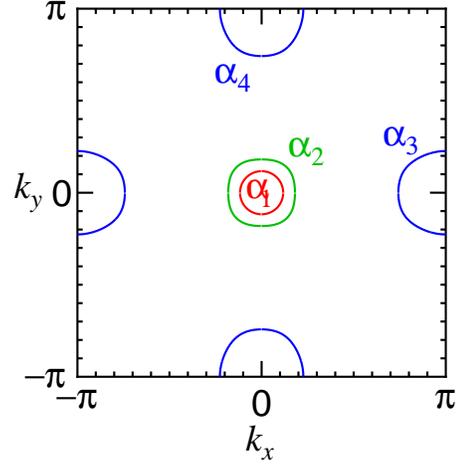}
	\caption{\label{fig:fs} Fermi surfaces of the 5 bands model for iron pnictide in the unfolded Brillouin zone.}
\end{figure}

 Now, we study the 10 $\times$ 10 Nambu BCS Hamiltonian $\hat{\mathcal{H}}_{\bm{k}}$ composed of the five orbital tight binding model\cite{kuroki_unconventional_2008} and the band-diagonal SC gap.
\cite{onari_violation_2009} 
 The Hamiltonian is given by
\begin{eqnarray}
	\hat{\mathcal{H}}^{0}
	&=&
	\sum_{\bm{k}}
	\hat{c}_{\bm{k}}^{\dagger}
	\hat{\mathcal{H}}_{\bm{k}}^{0}
	\hat{c}_{\bm{k}}
,
\end{eqnarray}
where $\hat{c}_{\bm{k}}^{\dagger}$ and $\hat{c}_{\bm{k}}$ are vectors, 
\begin{eqnarray}
	\hat{c}_{\bm{k}}^{\dagger}
	=
	\left(
		c_{1,\bm{k},\uparrow}^{\dagger},\cdots
		c_{5,\bm{k},\uparrow}^{\dagger},
		c_{1,- \bm{k},\downarrow},\cdots
		c_{5,- \bm{k},\downarrow},
	\right)
,
\end{eqnarray}
and $c_{\alpha,\bm{k},\sigma}^{\dagger}$ ($c_{\alpha,\bm{k},\sigma}$) is a creation (annihilation) operator of an electron for band $\alpha$ with wave vector $\bm{k}$ and spin $\sigma$.
$\hat{\mathcal{H}}_{\bm{k}}^{0}$ is 10 $\times$ 10 Nambu BCS Hamiltonian given as 
\begin{eqnarray}
	\hat{\mathcal{H}}_{\bm{k}}^{0}
	=
	\left(
	\begin{array}{cc}
		\hat{H}_{\bm{k}} & \hat{\Delta}_{\bm{k}} \\
		\hat{\Delta}_{\bm{k}}^{\dagger} & - \hat{H}_{\bm{k}}
	\end{array}
	\right)
,
\end{eqnarray}
where $\hat{H}_{\bm{k}}$ and $\hat{\Delta}_{\bm{k}}$ are 5 $\times$ 5 matrices in the band-diagonal basis,
\begin{eqnarray}
	\hat{H}_{\bm{k}}
	=
	\left(
	\begin{array}{ccc}
		\epsilon_{1,\bm{k}}	&	&	0	\\
			&	\ddots 	&	\\
		0	&	&	\epsilon_{5,\bm{k}}
	\end{array}
	\right)
,
% \nonumber
% \end{eqnarray}
\quad
% \begin{eqnarray}
	\hat{\Delta}_{\bm{k}}
	=
	\left(
	\begin{array}{ccc}
		\Delta_{1,\bm{k}}	&	&	0	\\
			&	\ddots 	&	\\
		0	&	&	\Delta_{5,\bm{k}}
	\end{array}
	\right)
,
\end{eqnarray}
where $\epsilon_{\alpha,\bm{k}}$ and $\Delta_{\alpha,\bm{k}}$ are a dispersion and a gap function of quasi particle for $\alpha$ and $\bm{k}$, respectively. 
 Figure \ref{fig:fs} shows the Fermi surface (1 Fe atom/unit cell) for electron doped iron pnictides. 
 Fermi surface $\alpha_{1}$ and $\alpha_{2}$ around $\Gamma$ point are hole pockets and $\alpha_{3}$ and $\alpha_{4}$ are electron pockets.
 Hereafter, we approximate $\Delta_{\alpha,\bm{k}}$ as follows:
\begin{eqnarray}\label{eq:delta}
	\Delta_{\alpha,\bm{k}}
	\approx
	\Delta_{\alpha}
	=
	\Delta_{\alpha}^{0}
	\tanh
	\left(
	\frac{\pi}{2}
	\sqrt{\frac{T_{\rm c}}{T}-1}
	\right)
,
\end{eqnarray}
where $\Delta_{\alpha}^{0}$ is a superconducting gap at zero temperature and, in the preset work, we assume that the superconducting gaps for each band are $\bm{k}$-independent. 

 The 10 $\times$ 10 Green's function $\hat{\mathcal{G}}_{\bm{k}}$ in the Nambu representation is given by \cite{allen_theory_1983}
\begin{eqnarray}\label{eq:green}
	\hat{\mathcal{G}}_{\bm{k}} ( i \omega_{n} )
% 	&=&
% 	\left(
% 	\begin{array}{cc}
% 		\hat{G}_{\bm{k}} ( i \omega_{n} )	&	 \hat{F}_{\bm{k}} ( i \omega_{n} ) \\
% 		\hat{F}_{\bm{k}} ( i \omega_{n} )	&	-\hat{G}_{\bm{k}} (- i \omega_{n} )
% 	\end{array}
% 	\right) \\
	&=&
	\left(
		i \omega_{n}\hat{1}
		- \hat{\mathcal{H}}_{\bm{k}}^{0}
		- \hat{\Sigma}_{\bm{k}} ( i \omega_{n} )
	\right)^{-1}
,
\end{eqnarray}
where $\omega_{n} = \pi T (2n + 1)$ is the fermion Matsubara frequency.
 The normal self energy $\hat{\Sigma}_{\bm{k}} ( i \omega_{n} )$ represents % the contribution to the particle's energy and effective mass
 the inelastic quasi-particle damping and mass enhancement due to many-body interactions. 
 We neglect the impurity induced self energy  since it does not change the density of states (DOS) in a simple $s$-wave state, Eq. (\ref{eq:delta}) (Anderson theorem). 
% So we do not consider the impurity scattering in the present study. 
 When $\hat{\Sigma}_{\bm{k}} ( i \omega_{n} )$ has band diagonal form, Eq. (\ref{eq:green}) also becomes a band diagonal form, and the Green's function for band $\alpha$ is given by a 2 $\times$ 2 matrix, 
\begin{eqnarray}
	&&
	\hat{\mathcal{G}}_{\alpha,\bm{k}} ( i \omega_{n} )
\nonumber \\
	&=&
	\left(
	\begin{array}{cc}
		i \omega_{n}
		- \epsilon_{\alpha, \bm{k}}
		- \Sigma_{\alpha, \bm{k}} ( i \omega_{n} )
	&	- \Delta_{\alpha}
	\\
		- \Delta_{\alpha}
	&	i \omega_{n}
		+ \epsilon_{\alpha, \bm{k}}
		+ \Sigma_{\alpha, \bm{k}} ( - i \omega_{n} )
\\
	\end{array}
	\right)^{-1}
\nonumber \\
	&=&
	\frac{1}{|{\hat{\mathcal{G}}_{\alpha,\bm{k}} ( i \omega_{n} )}^{-1}|}
\nonumber \\
	&&
	\times
	\left(
	\begin{array}{cc}
		i \omega_{n}
		+ \epsilon_{\alpha, \bm{k}}
		+ \Sigma_{\alpha, \bm{k}} ( - i \omega_{n} )
	&	\Delta_{\alpha}
	\\
		\Delta_{\alpha}
	&	i \omega_{n}
		- \epsilon_{\alpha, \bm{k}}
		- \Sigma_{\alpha, \bm{k}} ( i \omega_{n} )
	\end{array}
	\right)
\nonumber \\
	& \equiv &
	\left(
	\begin{array}{cc}
		G_{\alpha,\bm{k}} ( i \omega_{n} )
	&	F_{\alpha,\bm{k}} ( i \omega_{n} )
	\\
		F_{\alpha,\bm{k}} ( i \omega_{n} )
	&	- G_{\alpha,\bm{k}} ( - i \omega_{n} )
	\end{array}
	\right)
,
\end{eqnarray}
where $|{\hat{\mathcal{G}}_{\alpha,\bm{k}} ( i \omega_{n} )}^{-1}|$ is the determinant of the inverse matrix of the Green's function, and $G_{\alpha, \bm{k}}$, $F_{\alpha, \bm{k}}$ and $\Sigma_{\alpha, \bm{k}}$ are the normal Green's function, the anomalous Green's function and the normal self-energy for band $\alpha$ with wave vector $\bm{k}$, respectively. 
 The local Green's function $\hat{\bf g}_{\alpha}$ is obtained by the $\bm{k}$ summation; 
\begin{eqnarray}\label{eq:integ}
	\hat{\bf g}_{\alpha} ( i \omega_{n} )
	&=&
	\frac{1}{N}
	\sum_{\bm{k}}
	\hat{\mathcal{G}}_{\alpha, \bm{k}} ( i \omega_{n} )
\nonumber \\
	&=&
	\int_{-\infty}^{\infty}
	d \epsilon
	N ( \epsilon )
	\hat{\mathcal{G}}_{\alpha} ( \epsilon, i \omega_{n} )
,
\end{eqnarray}
where $N ( \epsilon )$ is the quasi-particle DOS. 
 In this $\epsilon$ integral, the dominant contribution comes from the explicit $\epsilon$ term of the denominator $|{\hat{\mathcal{G}}_{\alpha,\bm{k}} ( i \omega_{n} )}^{-1}|$. 
 By neglecting the $\epsilon$ dependence of $N ( \epsilon )$ and applying the infinite dimensional approximation for $\Sigma_{\alpha} (\epsilon, i \omega_{n})$ (see the following subsection \ref{ssec:2nd}), 
we obtain
\begin{eqnarray}
	\hat{\bf g}_{\alpha} ( i \omega_{n} )	
	& \approx &
	\frac{- \pi N (0)}
		 {	\sqrt{{\tilde{\omega}_{\alpha} ( i \omega_{n} )}^{2}
% 			+ \tilde{\Delta}_{\alpha}^{2}}}
			+ \Delta_{\alpha}^{2}}}
% \nonumber \\
% 	&&
% 	\times
	\left(
	\begin{array}{cc}
		i \tilde{\omega}_{\alpha} ( i \omega_{n} )
% 	&	  \tilde{\Delta}_{\alpha}
 	&	  \Delta_{\alpha}
	\\
% 		  \tilde{\Delta}_{\alpha}
		  \Delta_{\alpha}
	&	i \tilde{\omega}_{\alpha} ( i \omega_{n} )
	\end{array}
	\right)
\nonumber \\
	& \equiv &
	\left(
	\begin{array}{cc}
		g_{\alpha} ( i \omega_{n} )
	&	f_{\alpha} ( i \omega_{n} )
	\\
		f_{\alpha} ( i \omega_{n} )
	&	g_{\alpha} ( i \omega_{n} )
	\end{array}
	\right)
.
\end{eqnarray}
From the five orbital model, the total DOS per spin is $N(0) = 0.66$~eV$^{-1}$ at Fermi level. 
Then, analytic continuation ($i \omega_{n} \rightarrow \tilde{\omega} = \omega + i \delta$) yields the retarded Green's function as follows, 
\begin{eqnarray}
	g_{\alpha}^{\rm R} ( \omega )	
	&=&
	\frac{- \pi N (0) \tilde{\omega}_{\alpha}^{\rm R} ( \omega )}
		 {	\sqrt{- {\tilde{\omega}_{\alpha}^{\rm R} ( \omega )}^{2}
			+ {\Delta}_{\alpha}^{2}}}
,
\nonumber \\
	f_{\alpha}^{\rm R} ( \omega )	
	&=&
	\frac{- \pi N (0) \Delta_{\alpha}}
		 {	\sqrt{- {\tilde{\omega}_{\alpha}^{\rm R} ( \omega )}^{2}
			+ {\Delta}_{\alpha}^{2}}}
,
\end{eqnarray}
where $\tilde{\omega}_{\alpha}^{\rm R} ( \omega )$ is defined as
\begin{eqnarray}
	\tilde{\omega}_{\alpha}^{\rm R} ( \omega )
	&=&
	\tilde{\omega}
	- i {\rm Im} \Sigma_{\alpha}^{\rm R} ( \omega ) 
	\nonumber
	\\
	&=&
	\tilde{\omega}
	+ i \gamma_{\alpha}^{*} ( \omega )
,
\end{eqnarray}
where $\gamma^{*}$ is the ``renormalized" quasi-particle damping, which is described by using ``unrenormalized" quasi-particle damping $\gamma \equiv - {\rm Im} \Sigma^{\rm R}$ and bare and effective masses $m$ and $m^{*}$ as 
\begin{eqnarray}\label{eq:gamma_mass}
	\gamma^{*}
	\equiv
	\frac{m}{m^{*}} \gamma
.
\end{eqnarray}
 The mass enhancement factor $m^{*}/m$ is given by
\begin{eqnarray}
	\frac{m^{*}}{m}
	=
	1 - \lim_{\omega \rightarrow 0} \frac{\partial {\rm Re} \Sigma^{\rm R} ( \omega )}{\partial \omega}
,
\end{eqnarray}
which is reported as $1 \sim 3$ for iron pnictides called 1111 and 122 systems from various experiments such as de Haas-van Alphen measurements, \cite{sebastian_quantum_2008, analytis_fermi_2009, shishido_evolution_2010} optical spectral weight, \cite{qazilbash_electronic_2009} and Seebeck effect with specific heat. \cite{okuda_thermoelectric_2011}
 The mass enhancement factor has a relatively large value in the optimally doped systems and gradually decreases by carrier doping.

%%%%%%%%%%%%%%%%%%%%%%%%%%%%%%%%%%%%%%%%%%%%%%%%%%%%%%%%%%%%%%%%%%%%%%%%%%%%%%%
\subsection{Nuclear Magnetic Relaxation Rate}
 The nuclear magnetic relaxation rate $1/T_{1}T$ in the superconducting state is given by the standard formula:
\begin{eqnarray}\label{eq:t1t-def}
	\frac{1}{T_{1}T}
	& \propto &
	\frac{1}{N}
	\sum_{\alpha, \bm{q}}
	\lim_{\omega \rightarrow 0}
	{\rm Im}
	\frac{\chi_{\alpha, \bm{q}}^{\rm R} ( \omega )}{\omega}
,
\end{eqnarray}
where $\chi_{\bm{q}}^{\rm R} ( \omega )$ is the superconducting spin susceptibility and given by 
\begin{eqnarray}\label{eq:chi}
	{\rm Im} \chi_{\alpha, \bm{q}}^{\rm R} ( \omega )
	&=&
	\frac{1}{2 \pi N}
	\sum_{\beta, \bm{k}}
	\int_{-\infty}^{\infty} d \omega'
	\left(
		\tanh \frac{\omega + \omega'}{2T} 
		-
		\tanh \frac{\omega'}{2T} 
	\right)
\nonumber \\
	&&
	\times
	\left(
		{\rm Im} G_{\beta, \bm{k}+\bm{q}}^{\rm R} ( \omega + \omega' )
		{\rm Im} G_{\alpha, \bm{k}}^{\rm R} ( \omega' )
	\right.
\nonumber \\
	&& \quad
	\left.
		+
		{\rm Im} F_{\beta, \bm{k}+\bm{q}}^{\rm R} ( \omega + \omega' )
		{\rm Im} F_{\alpha, \bm{k}}^{\rm R} ( \omega' )
	\right)
.
\end{eqnarray}
 In Eq. (\ref{eq:t1t-def}), the limitation becomes a differential of hyperbolic tangent and each summation of $\alpha$, $\beta$, $\bm{k}$ and $\bm{q}$ can be calculated independently.
 Then, $1/T_{1}T$ becomes 
\begin{eqnarray}\label{eq:t1t}
	\frac{1}{T_{1}T}
% 	&=&
% 	\lim_{\omega \rightarrow 0}
% 	{\rm Im}
% 	\frac{\chi_{\alpha, \bm{q}}^{\rm R} ( \omega )}{\omega}
% \nonumber \\
	& \propto &
	\int_{-\infty}^{\infty} d \omega
	\frac{{{\rm Im} g^{\rm R} ( \omega )}^2
		+
		{{\rm Im} f^{\rm R} ( \omega )}^2}{4 \pi T\cosh^{2} ( \omega / 2T )}
% \nonumber \\
% 	&&
% 	\times
% 	\left(
% 		{{\rm Im} G^{\rm R} ( \omega )}^2
% 		+
% 		{{\rm Im} F^{\rm R} ( \omega )}^2
% 	\right)
\nonumber \\
	& \equiv &
	\int_{-\infty}^{\infty} d \omega
	X ( \omega )
,
\end{eqnarray}
with
\begin{eqnarray}
	g^{\rm R} ( \omega )
	& \equiv &
	\sum_{\alpha}
	g_{\alpha}^{\rm R} ( \omega )
,
\nonumber \\
	f^{\rm R} ( \omega )
	& \equiv &
	\sum_{\alpha}
	f_{\alpha}^{\rm R} ( \omega )
,
\end{eqnarray}
where the integrand of $\omega$ was defined as $X ( \omega )$.

%%%%%%%%%%%%%%%%%%%%%%%%%%%%%%%%%%%%%%%%%%%%%%%%%%%%%%%%%%%%%%%%%%%%%%%%%%%%%%%
\subsection{Second order perturbation theory}\label{ssec:2nd}
 A quasi-particle excitation is well defined if the damping rate $\gamma$ is small compared to the energy scale. 
 In this case, the quasi-particle damping rate $\gamma$ can generally be computed from the imaginary part of the normal self energy, 
\cite{weldon_simple_1983} 
\begin{eqnarray}\label{eq:gamma-def}
	{\rm Im} \Sigma_{\bm{k}}^{\rm R} ( \omega ) = - \gamma_{\bm{k}} ( \omega )
.
\end{eqnarray}

 By using the second order perturbation theory, we derive the microscopic picture of $\gamma$ and analyze the NMR experimental result. 
 We calculate the electron-electron scattering based on the single band Hubbard model for simplicity. 
 From the second order perturbation theory, the self energy $\Sigma_{\bm{k}}$ due to the electron-electron scattering is given by 
\begin{eqnarray}
	\Sigma_{\bm{k}} ( i \omega_{n} )
	=
	\frac{V_{\rm eff}^{2} T}{N}
	\sum_{\bm{k}', n'}
	G_{\bm{k}'}^{0} ( i \omega_{n'} )
	\chi_{\bm{k}-\bm{k}'}^{0} ( i \omega_{n} - i \omega_{n'} )
,
\end{eqnarray}
where $V_{\rm eff}$ is the effective electron-electron interaction enhanced by spin and orbital fluctuations. 
 Its value can be estimated from the observed conductivity in subsection \ref{ssec:damp}. 
 The imaginary part of the retarded self energy, which is obtained by the analytic continuation, is given by,
\begin{eqnarray}\label{eq:2nd}
	{\rm Im} \Sigma_{\bm{k}}^{\rm R} ( \omega )
	&=&
	\frac{V_{\rm eff}^{2}}{2 \pi N}
	\sum_{\bm{k}'}
	\int_{-\infty}^{\infty} d \omega'
	\left(
		\tanh \frac{\omega'}{2T}
		+
		\coth \frac{\omega - \omega'}{2T}
	\right)
\nonumber \\
	&&
	\times
	{\rm Im} G_{\bm{k}'}^{0, \rm R} ( \omega' )
	{\rm Im} \chi_{\bm{k}-\bm{k}'}^{0, \rm R} ( \omega - \omega' )
,
\end{eqnarray}
where the superscript ``0" on the Green function and susceptibility means %unperturbed state. 
the absence of self-energy correction. 

 In the infinite dimensional approximation, the each summation of wave vectors can be taken independently and we 
obtain the quasi-particle damping rate due to the electron-electron scattering as,
\begin{eqnarray}\label{eq:gamma}
	\gamma ( \omega ) 
	& \equiv &
	\frac{1}{N} \sum_{\bm{k}} \gamma_{\bm{k}} ( \omega )
	=
	- \frac{1}{N} \sum_{\bm{k}} {\rm Im} \Sigma_{\bm{k}} ( \omega )
\nonumber \\
	&=&
	- \frac{V_{\rm eff}^{2}}{2 \pi}
	\int_{-\infty}^{\infty} d \omega'
	\left(
		\tanh \frac{\omega'}{2T}
		+
		\coth \frac{\omega - \omega'}{2T}
	\right)
\nonumber \\
	&&
	\times
	{\rm Im} g^{0, \rm R} ( \omega' )
	{\rm Im} \chi^{0, \rm R} ( \omega - \omega' )
,
\end{eqnarray}
with bare susceptibility, 
\begin{eqnarray}\label{eq:chi0}
	{\rm Im} \chi^{0, \rm R} ( \omega )
	&=&
	\frac{1}{2 \pi}
	\int_{-\infty}^{\infty} d \omega'
	\left(
		\tanh \frac{\omega + \omega'}{2T} 
		-
		\tanh \frac{\omega'}{2T} 
	\right)
\nonumber \\
	&&
	\times
	\left(
		{\rm Im} g^{0, \rm R} ( \omega + \omega' )
		{\rm Im} g^{0, \rm R} ( \omega' )
	\right.
\nonumber \\
	&& \quad
	\left.
		+
		{\rm Im} f^{0, \rm R} ( \omega + \omega' )
		{\rm Im} f^{0, \rm R} ( \omega' )
	\right)
.
\end{eqnarray}

\begin{widetext}
To see the role of coherence factor,
we rewrite the imaginary part of the Green's functions 
with ${\rm Im} G_{\bm{k}}^{0, \rm R} ( \omega ) = - \frac{\pi}{2} \{ (1 + \frac{\epsilon_{\bm{k}}}{E_{\bm{k}}}) \delta (\omega - E_{\bm{k}}) + (1 - \frac{\epsilon_{\bm{k}}}{E_{\bm{k}}}) \delta (\omega + E_{\bm{k}}) \}$ and ${\rm Im} F_{\bm{k}}^{0, \rm R} ( \omega ) = - \frac{\pi \Delta_{\bm{k}}}{2 E_{\bm{k}}} \{ \delta (\omega - E_{\bm{k}}) - \delta (\omega + E_{\bm{k}}) \}$, where $E_{\bm{k}} \equiv \sqrt{\epsilon_{\bm{k}}^{2} + \Delta_{\bm{k}}^{2}}$.
By neglecting the energy-dependence of the DOS near the Fermi level,
we obtain another notation of the self energy as, 
\begin{eqnarray}
	{\rm Im} \Sigma_{\bm{k}}^{\rm R} ( \omega )
	&=&
	- \frac{\pi V_{\rm eff}^{2}}{4 N^{2}}
	\sum_{\bm{k}'}
	\sum_{\bm{k}''}
	\frac{\cosh \cfrac{\omega}{2 T}}
		 {\cosh \cfrac{E_{\bm{k}'}}{2 T} \cosh \cfrac{E_{\bm{k}-\bm{k}'+\bm{k}''}}{2 T} \cosh \cfrac{E_{\bm{k}''}}{2 T}}
\nonumber \\
	&& \times
	\frac{1}{8}
 	\biggl[
	\left(
		1 + \frac{\Delta_{\bm{k}-\bm{k}'+\bm{k}''} \Delta_{\bm{k}''}}{E_{\bm{k}-\bm{k}'+\bm{k}''} E_{\bm{k}''}}
 	\right)
	\big\{
		\delta ( |\omega| - |E_{\bm{k}'} + E_{\bm{k}-\bm{k}'+\bm{k}''} - E_{\bm{k}''}| )
%		+
%		\delta ( \omega + E_{\bm{k}'} + E_{\bm{k}-\bm{k}'+\bm{k}''} - E_{\bm{k}''} )
%\nonumber \\
%	&& \qquad \qquad \qquad \qquad \qquad \qquad
		+
		\delta ( |\omega| - |E_{\bm{k}'} - E_{\bm{k}-\bm{k}'+\bm{k}''} + E_{\bm{k}''}| )
%		+
%		\delta ( \omega + E_{\bm{k}'} - E_{\bm{k}-\bm{k}'+\bm{k}''} + E_{\bm{k}''} )
	\big\}
\nonumber \\
	&& \quad
	+
	\left(
		1 - \frac{\Delta_{\bm{k}-\bm{k}'+\bm{k}''} \Delta_{\bm{k}''}}{E_{\bm{k}-\bm{k}'+\bm{k}''} E_{\bm{k}''}}
	\right)
	\big\{
		\delta ( |\omega| - |E_{\bm{k}'} - E_{\bm{k}-\bm{k}'+\bm{k}''} - E_{\bm{k}''}| )
%		+
%		\delta ( \omega + E_{\bm{k}'} - E_{\bm{k}-\bm{k}'+\bm{k}''} - E_{\bm{k}''} )
%\nonumber \\
%	&& \qquad \qquad \qquad \qquad \qquad \qquad
		+
		\delta ( |\omega| - |E_{\bm{k}'} + E_{\bm{k}-\bm{k}'+\bm{k}''} + E_{\bm{k}''}| )
%		+
%		\delta ( \omega + E_{\bm{k}'} + E_{\bm{k}-\bm{k}'+\bm{k}''} + E_{\bm{k}''} )
	\big\}
	\biggr]
.
\end{eqnarray}
% by using 
% \begin{eqnarray}
% 	{\rm Im} G_{\bm{k}}^{0, \rm R}
% 	&=&
% 	{\rm Im} \frac{\omega}{ (\omega + E_{\bm{k}} + i \delta )(\omega - E_{\bm{k}} + i \delta )}
% \nonumber \\
% 	&=&
% 	\frac{1}{2}
% 	\left(
% 		\delta ( \omega - E_{\bm{k}} ) + \delta ( \omega + E_{\bm{k}} ) 
% 	\right)
% ,
% \end{eqnarray}
% and 
% \begin{eqnarray}
% 	{\rm Im} F_{\bm{k}}^{0, \rm R}
% 	&=&
% 	{\rm Im} \frac{\Delta}{ (\omega + E_{\bm{k}} + i \delta )(\omega - E_{\bm{k}} + i \delta )}
% \nonumber \\
% 	&=&
% 	\frac{\Delta}{2 E_{\bm{k}}} 
% 	\left(
% 		\delta ( \omega - E_{\bm{k}} ) - \delta ( \omega + E_{\bm{k}} ) 
% 	\right)
% ,
% \end{eqnarray}
%and for $\omega = 0$, 
%\begin{eqnarray}
%	{\rm Im} \Sigma_{\bm{k}}^{\rm R} ( 0 )
%	&=&
%	- \frac{\pi V_{\rm eff}^{2}}{4 N^{2}}
%	\sum_{\bm{k}'}
%	\sum_{\bm{k}''}
%	\frac{1}{\cosh \cfrac{E_{\bm{k}'}}{2 T} \cosh \cfrac{E_{\bm{k}-\bm{k}'+\bm{k}''}}{2 T} \cosh \cfrac{E_{\bm{k}''}}{2 T}}
%\nonumber \\
%	&& \times
%	\frac{1}{4}
% 	\biggl[
%	\left(
%		1 + \frac{\Delta_{\bm{k}-\bm{k}'+\bm{k}''} \Delta_{\bm{k}''}}{E_{\bm{k}-\bm{k}'+\bm{k}''} E_{\bm{k}''}}
% 	\right)
%	\big\{
%		\delta ( E_{\bm{k}'} + E_{\bm{k}-\bm{k}'+\bm{k}''} - E_{\bm{k}''} )
%		+
%		\delta ( E_{\bm{k}'} - E_{\bm{k}-\bm{k}'+\bm{k}''} + E_{\bm{k}''} )
%	\big\}
%\nonumber \\
%	&& \quad
%	+
%	\left(
%		1 - \frac{\Delta_{\bm{k}-\bm{k}'+\bm{k}''} \Delta_{\bm{k}''}}{E_{\bm{k}-\bm{k}'+\bm{k}''} E_{\bm{k}''}}
%	\right)
%		\delta ( E_{\bm{k}'} - E_{\bm{k}-\bm{k}'+\bm{k}''} - E_{\bm{k}''} )
%	\biggr]
%.
%\end{eqnarray}
For $T\lesssim T_{\rm c}$, the first term with the coherence factor
$(1+\Delta\Delta'/EE')$ gives the dominant contribution for the $s_{++}$-wave state.
At $T=0$, because of the thermal factor, the self energy becomes 
\begin{eqnarray}
	{\rm Im} \Sigma_{\bm{k}}^{\rm R} ( \omega )
	& \approx &
	- \frac{\pi V_{\rm eff}^{2}}{8 N^{2}}
	\sum_{\bm{k}'}
	\sum_{\bm{k}''}
	\left(
		1 - \frac{\Delta_{\bm{k}-\bm{k}'+\bm{k}''} \Delta_{\bm{k}''}}{E_{\bm{k}-\bm{k}'+\bm{k}''} E_{\bm{k}''}}
	\right)
	\delta ( |\omega| - |E_{\bm{k}'} + E_{\bm{k}-\bm{k}'+\bm{k}''} + E_{\bm{k}''}| )
.
\end{eqnarray}
It is clear from the delta function that it equals 0 for $\omega < 3\Delta_{\rm min}$. 
\end{widetext}

 Experimental studies on the iron pnictides such as ARPES \cite{richard_fe-based_2011}, point contact tunneling, \cite{samuely_possible_2009} NQR \cite{kawasaki_two_2008} and specific heat \cite{hardy_doping_2010} have demonstrated that there are at least two different superconducting gaps, small gap $\Delta_{\rm S}$ and large gap $\Delta_{\rm L}$. 
 Multi-gap superconductivity has been seen in a number of systems including MgB$_{2}$ which is an $s$-wave superconductor with a $T_{\rm c}$ of 39~K. \cite{nagamatsu_superconductivity_2001, liu_beyond_2001}
 Based on these experiments, we set the small and large superconducting gaps as,
\begin{eqnarray}
	&& \Delta_{\alpha_{1}}^{0}=\Delta_{\alpha_{2}}^{0}=\Delta_{\rm S}
,
\nonumber \\
	&& \Delta_{\alpha_{3}}^{0}=\Delta_{\alpha_{4}}^{0}=\pm \Delta_{\rm L}
,
\\
	&& \mbox{($+$: for $s_{++}$, $-$: for $s_{\pm}$)}
.
\nonumber
\end{eqnarray}
Hereafter, we employ these values as $2 \Delta_{\rm L} / T_{\rm c} = 5$ and $\Delta_{\rm L}/\Delta_{\rm S} = 3$, which are approximately satisfied in various Co-doped 122 systems according to the specific heat measurements. \cite{johnston_puzzle_2010}

\begin{figure}[tb]
 	\includegraphics[width=4cm]{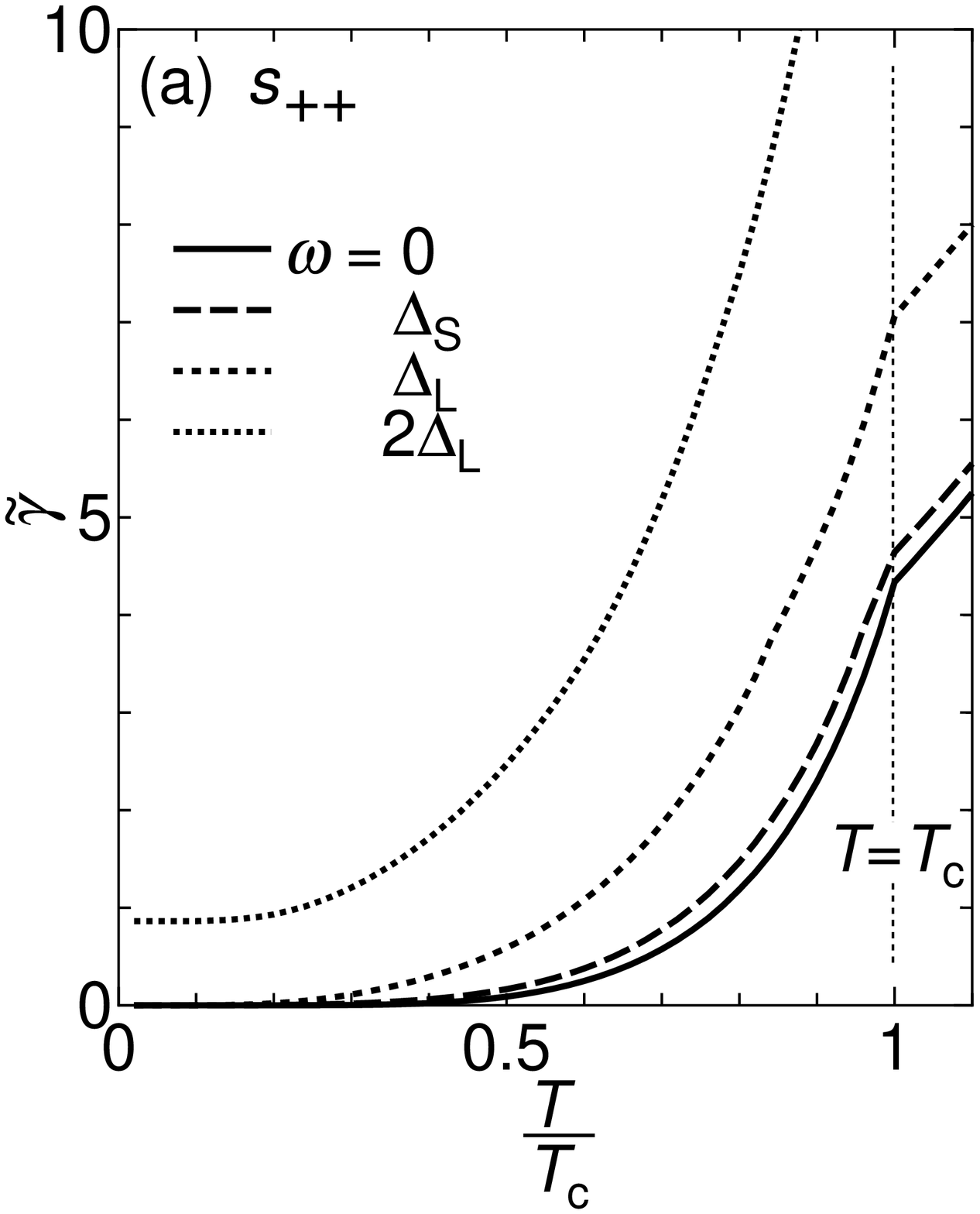}
 	\includegraphics[width=4cm]{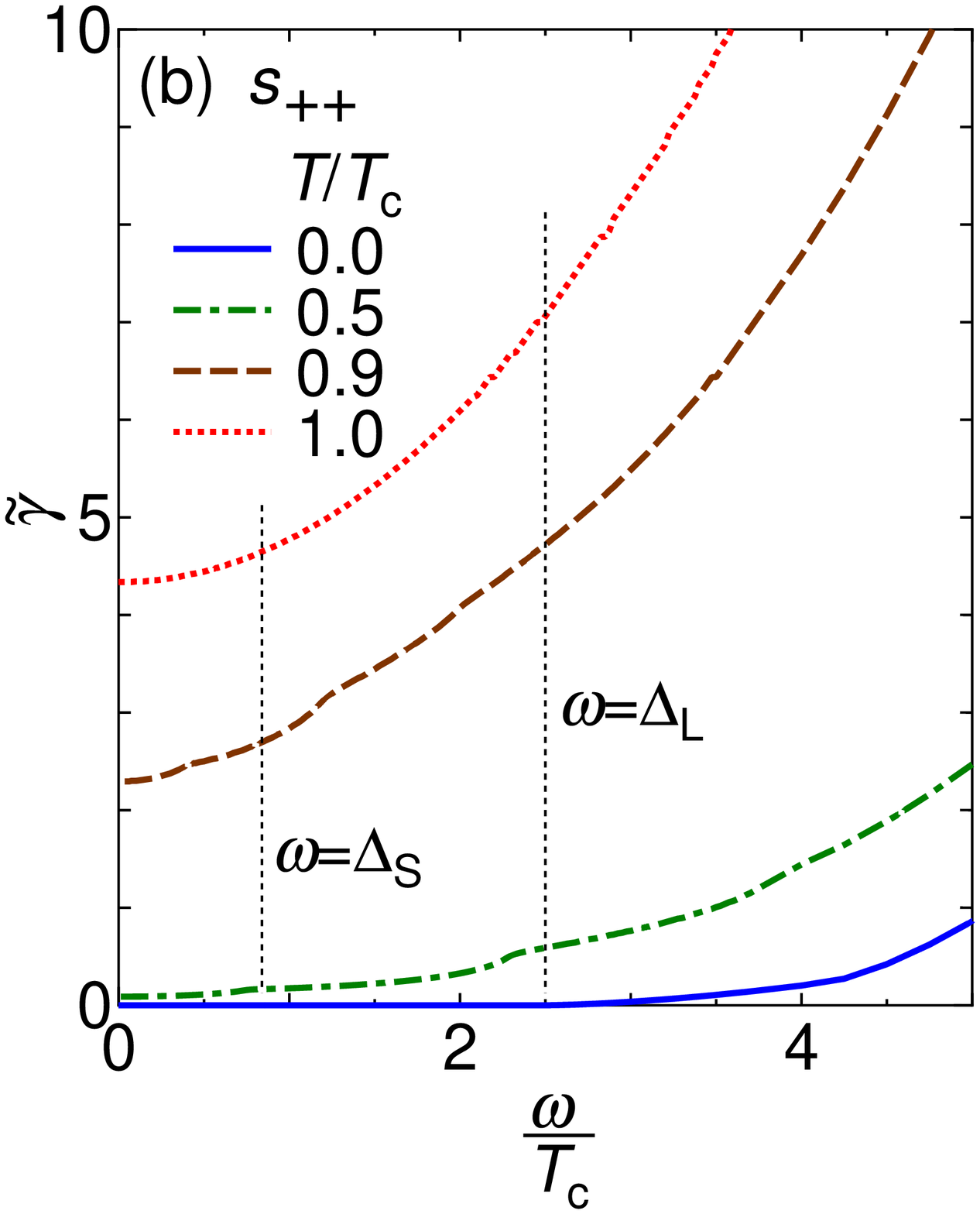}
	\caption{\label{fig:gx++}
(a) The temperature dependences of $\tilde{\gamma} \equiv \gamma/V_{\rm eff}^2 T_{\rm c}^2$ in the $s_{++}$-wave state with $2 \Delta_{\rm L}/T_{\rm c}=5$ and $\Delta_{\rm L}/\Delta_{\rm S}=3$. 
Solid, dashed, short-dashed and dotted lines corresponds to $\omega=0$, $\Delta_{\rm S}$, $\Delta_{\rm L}$ and $2\Delta_{\rm L}$, respectively. 
(b) $\omega$ dependences of $\tilde{\gamma}$ in the $s_{++}$-wave state for various $T$. 
Solid, dashed-dotted, dashed and dotted lines represent $T = 0$, $0.5T_{\rm c}$, $0.9T_{\rm c}$ and $T_{\rm c}$, respectively.
Vertical dotted lines show $\omega = \Delta_{\rm L}$ and $\Delta_{\rm S}$. 
}
\end{figure}

 Figure \ref{fig:gx++} (a) shows the temperature dependence of $\tilde{\gamma} \equiv \gamma/V_{\rm eff}^2 T_{\rm c}^2$ on the $s_{++}$-wave superconducting state at various frequencies. 
 The $T$ dependence of $\tilde{\gamma}$ is small in the normal state ($T>T_{\rm c}$), but $\tilde{\gamma}$ rapidly decreases with decreasing $T$ in the superconducting state ($T<T_{\rm c}$). 
 Figure \ref{fig:gx++} (b) shows the $\omega$ dependence of $\tilde{\gamma}$.  
 Two vertical dotted lines show the small and large superconducting gaps at zero temperature $\Delta_{\rm S}$ and $\Delta_{\rm L}$, respectively. 
 In the normal state, $\tilde{\gamma}$ is large due to the strong correlation. 
 However, $\tilde{\gamma}$ is strongly suppressed in the superconducting state since the inelastic damping $\tilde{\gamma}$ is reduced as the superconducting gap opens. 
 It is satisfied that $\gamma ( \omega ) |_{T = 0} = 0$ for $\omega < 3 \Delta_{\rm S} = \Delta_{\rm L}$. 

\begin{figure}[tb]
	\includegraphics[width=4cm]{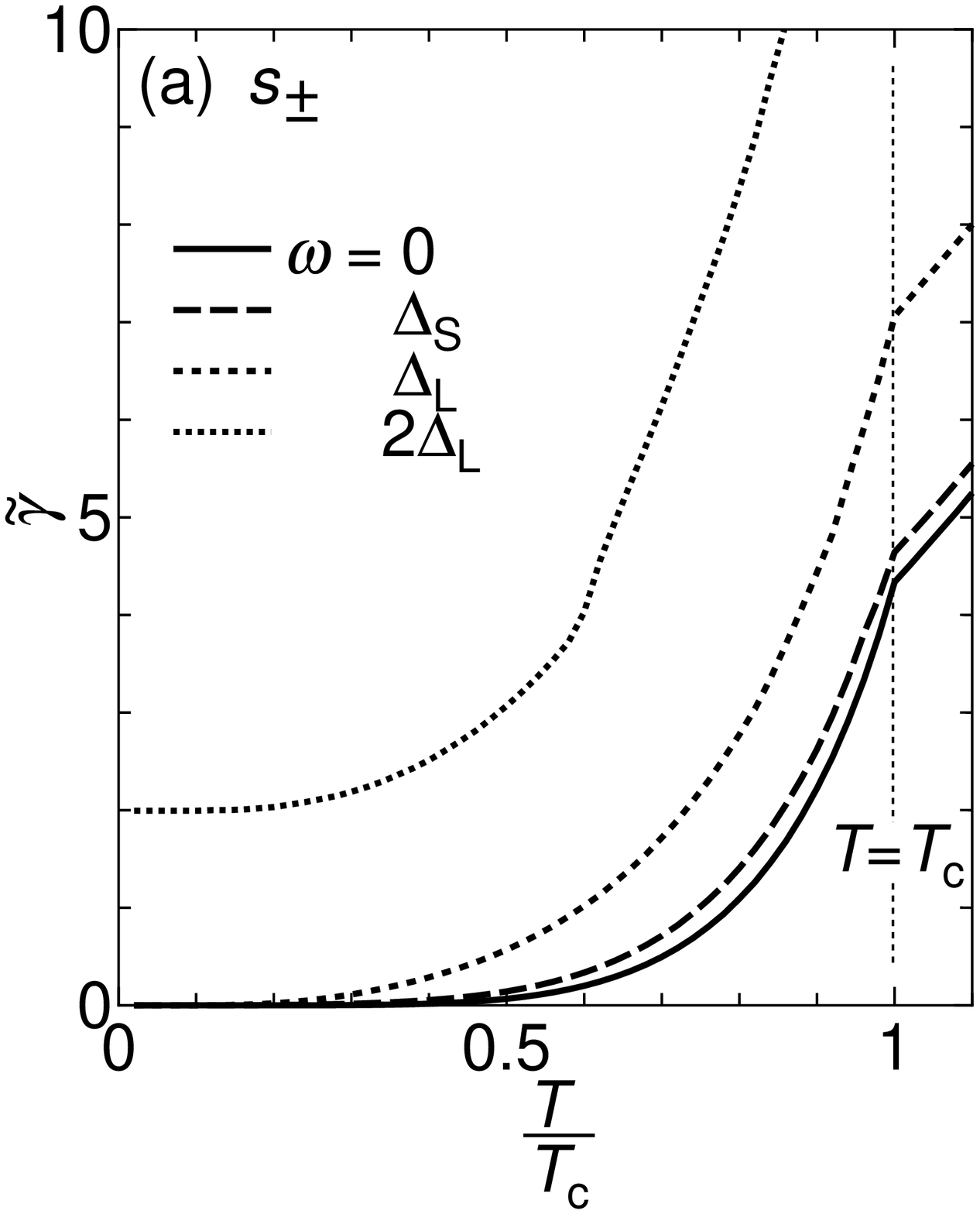}
	\includegraphics[width=4cm]{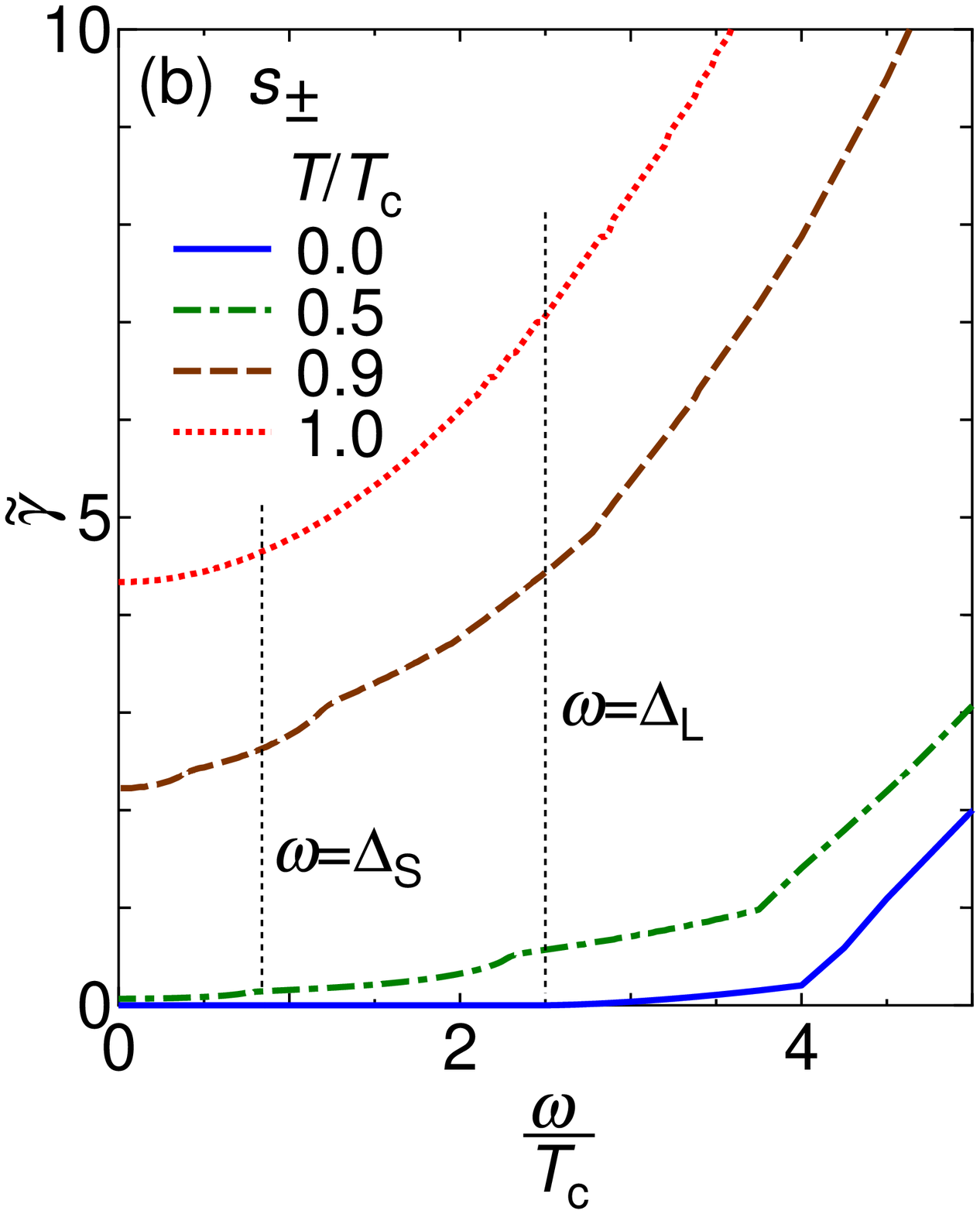}
	\caption{\label{fig:gx+-} 
(a) The $T$ dependences of the $\tilde{\gamma}$ in the $s_{\pm}$-wave state with $2 \Delta_{\rm L}/T_{\rm c}=5$ and $\Delta_{\rm L}/\Delta_{\rm S}=3$. 
(b) The $\omega$ dependences of the $\tilde{\gamma}$ in the $s_{\pm}$-wave state. 
}
\end{figure}

 Figures \ref{fig:gx+-} (a) and (b) show the $T$ and $\omega$ dependence of $\tilde{\gamma}$ on the $s_{\pm}$-wave superconducting state, respectively. 
 As is the case in $s_{++}$, $\tilde{\gamma}$ is strongly suppressed for $T<T_{\rm c}$. 
 The obtained quasi-particle damping for $s_{\pm}$-wave state is similar to the $s_{++}$-wave state.

%%%%%%%%%%%%%%%%%%%%%%%%%%%%%%%%%%%%%%%%%%%%%%%%%%%%%%%%%%%%%%%%%%%%%%%%%%%%%%%
\subsection{Quasiparticle Damping}\label{ssec:damp}
 In this subsection, we estimate the absolute value of the inelastic damping rate $\gamma (0)$ in the normal state, and thus yields the effective interaction $V_{\rm eff}$, from the experimentally observed resistivity. 
 From the Nakano-Kubo formula, the conductivity is given by, 
\begin{eqnarray}
	\sigma
	=
	\frac{e^{2}}{4 \pi c}
	\sum_{\alpha}
	\int_{\rm FS \alpha} d k_{\parallel}
	\frac{|v_{\alpha, \bm{k}}|}{2 \gamma ( 0 )}
,	
\end{eqnarray}
where $\gamma (0)$ is the ``unrenormalized" damping at zero energy, and $v_{\alpha, \bm{k}}$ is the Fermi velocity at $\bm{k}$ on the $\alpha$ th Fermi surface. 
Here, we neglect the current vertex correction 
since it is not important for the diagonal conductivity \cite{kontani-ROP}.
 $\gamma (0)$ is derived from the theoretical relation between $\rho(T)$ and $\gamma (0)$, \cite{onari_neutron_2011}
\begin{eqnarray}\label{eq:rt-r0}
	\rho (T) - \rho (0)
	\approx
	\begin{cases}
		0.0020 \gamma (0) \text{ [$\mu \Omega$cm]}	&	\text{for $c=6$~\AA} \\
		0.0028 \gamma (0) \text{ [$\mu \Omega$cm]}	&	\text{for $c=8$~\AA}
	\end{cases}
.
\end{eqnarray}
 The interlayer spacing $c \approx 6$~\AA{} and $c \approx 8$~\AA{} are corresponding to the 122 and 1111 systems, respectively. 
 On the other hand, we solve Eq. (\ref{eq:gamma}) as a normal state (${\rm Im} f^{0, \rm R} ( \omega ) = 0$) and the inelastic damping rate $\gamma(0)$ is represented as
\begin{eqnarray}\label{eq:gamma0}
	{\gamma (0)}|_{T=T_{\rm c}}
	=
	\frac{\pi^{3}}{2}
	V_{\rm eff}^2
	{N (0)}^{3}
	T_{\rm c}^{2}
.
\end{eqnarray}
 Then, the effective interaction $V_{\rm eff}$ is derived by comparing Eq. (\ref{eq:rt-r0}) with Eq. (\ref{eq:gamma0}).

 In Table \ref{tb:gamma}, we show $T_{\rm c}$ and other parameters of Ba(Fe$_{1-x}$Co$_{x}$)$_{2}$As$_{2}$ for various doping rate $x$. 
 The resistivity due to inelastic scattering, $\rho (T) - \rho (0)$, is estimated by fitting the experimental data below 150~K \cite{rullier-albenque_hall_2009} with
\begin{eqnarray}
	\rho (T) = \rho (0) + a T^{n}
.
\end{eqnarray}
where $\rho (0)$, $a$ and $n$ are free parameters. 
 The obtained $\gamma = 69$~K for the optimally doped compound ($x=0.08$) is larger than the transition temperature $T_{\rm c} = 24$~K. 
 On the other hand, $\gamma$ decreases rapidly by carrier doping, and $\gamma = 7.5$~K becomes smaller than $T_{\rm c} = 11$~K for the over doped compound ($x=0.14$). 
 The estimated effective interaction $V_{\rm eff}$ is 17.7~eV for the optimally doped system, and 12.2~eV for the over doped system. 
 $V_{\rm eff}$ is large as compared with the bare on-site Coulomb interaction $U = 2 \sim 3$~eV for $3d$ electrons on a Fe atom \cite{miyake_comparison_2010} because of the spin and charge Stoner enhancement that give large spin and orbital fluctuations.

\begin{table}[tb]
\caption{\label{tb:gamma}
 $T_{\rm c}$ and other parameters of Ba(Fe$_{1-x}$Co$_{x}$)$_{2}$As$_{2}$ for various doping rate $x$. \cite{rullier-albenque_hall_2009}
 $\gamma(0)$ is estimated by fitting the experimental data and $V_{\rm eff}$ is obtained by $\gamma(0)$ and $N(0) = 0.66$~eV$^{-1}$. 
 We note that we obtain $n=1.0$ and $a=0.63$ ${\rm \mu \Omega cm / K}$ for $x=0.08$ according to the measurement by Sefat {\it et al.}
 \cite{Sefat}
 In this case, we obtain $V_{\rm eff}= 19.9$~eV.
 In this table, the units of $\rho$ and a are ${\rm \mu \Omega cm}$ and ${\rm \mu \Omega cm / K^{n}}$, respectively. 
 }
\begin{tabular*}{8cm}{@{\extracolsep{\fill}}c|cccc} \hline \hline
 	 \multicolumn{5}{c}{Ba(Fe$_{1-x}$Co$_{x}$)$_{2}$As$_{2}$
	\cite{rullier-albenque_hall_2009}} % & \multicolumn{2}{c}{LaFeAsO$_{1-x}$F$_{x}$} \cite{arushanov_resistivity_2010}
 	\\ \hline
	$x$
% 	& 0.07 
	& 0.08 & 0.12 & 0.14 & 0.20 % & 0.1 & 0.2
	\\ \hline
	$T_{\rm c}$
% 	& 25 
	& 24~K & 17~K & 11~K & -  % & 26 & 9
	\\
	$\rho(0)$
% 	& 89 
	& 73 & 65 & 65 & 57 % & 53(160) & 33(100)
	\\
	$\rho(T_{\rm c}) - \rho(0)$
% 	& 8.9 
	& 11.9 & 4.6 & 1.3 & - % & 6.6(20) & 0.57(17)
	\\
	$a$
% 	& 0.22 
	& 0.495 & 0.153 & 0.045 & 0.0011 % & (0.00295) & (0.0021)
	\\
	$n$
% 	& 1.15 
	& 1.0 & 1.2 & 1.4 & 2.0 % & 2.0 & 2.0
	\\
	$\gamma(0)|_{T=T_{\rm c}}$
% 	& 0.0045 
	& 0.0059 & 0.0023 & 0.00065 & - % & 0.0023 & 0.00020
	\\
	& (= 69~K) & (= 27~K) & (= 7.5~K) & -
	\\
	$V_{\rm eff}$
% 	& 14.7 
	& 17.7 & 15.5 & 12.7	& 4.3 % & 10.3 & 8.7
	\\
	\hline \hline
\end{tabular*}
\end{table}

%%%%%%%%%%%%%%%%%%%%%%%%%%%%%%%%%%%%%%%%%%%%%%%%%%%%%%%%%%%%%%%%%%%%%%%%%%%%%%%
\section{Results}\label{sec:results}

\begin{figure}[tb]
 	\includegraphics[width=6cm]{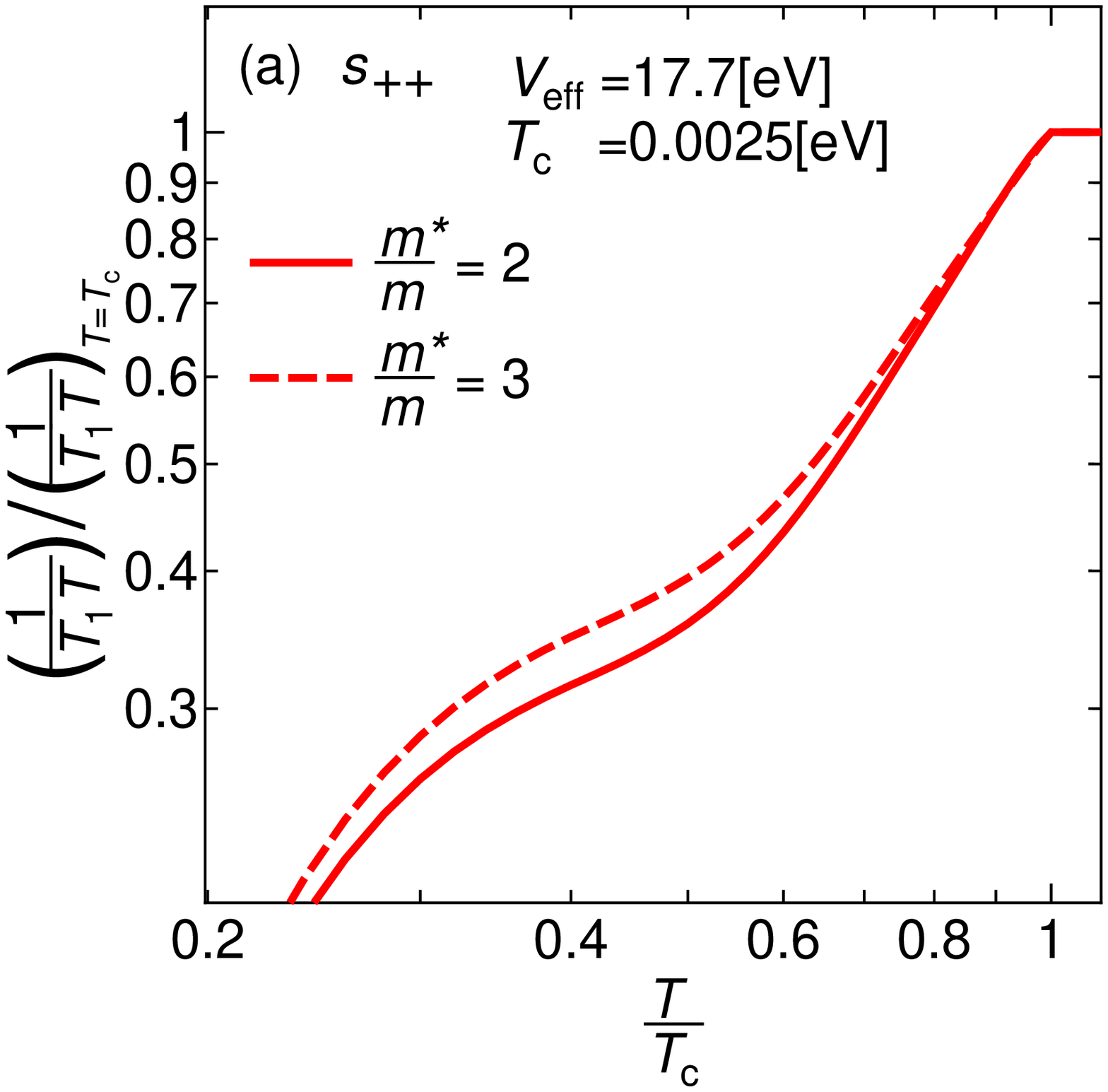}
 	\includegraphics[width=6cm]{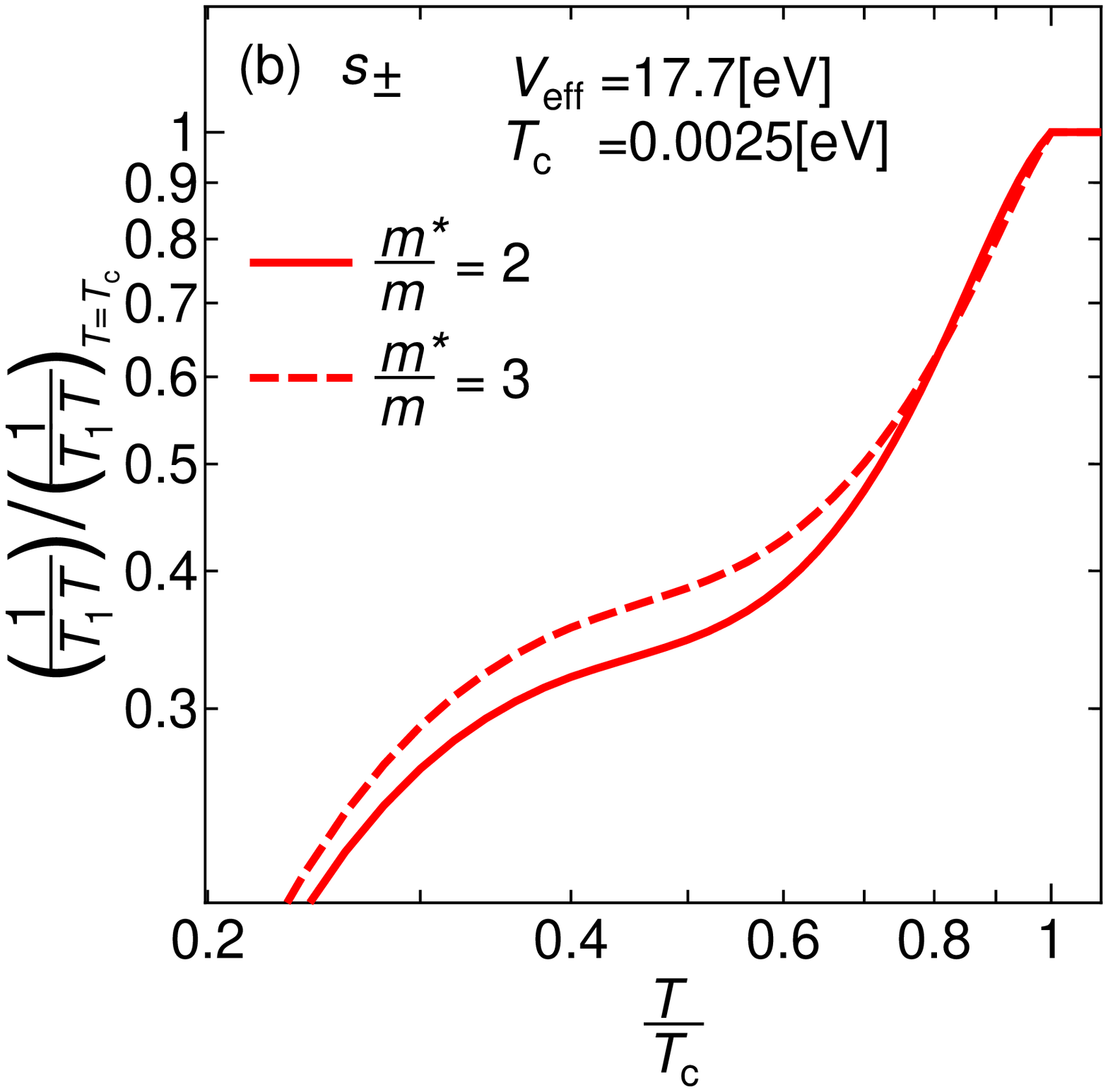}
	\caption{\label{fig:t1t_opt} The normalized nuclear relaxation rate $(1/T_{1}T)/(1/T_{1}T)|_{T=T_{\rm c}}$ for optimally doping state ($T_{\rm c} = 0.0025$~eV and $V_{\rm eff} = 17.7$~eV) in the (a) $s_{++}$ and (b) $s_{\pm}$-wave states as a function of reduced temperature $T/T_{\rm c}$. 
	Solid and dashed lines represent $m^{*}/m = 2$ and $m^{*}/m = 3$, respectively. 
	}
\end{figure}

\begin{figure}[tb]
 	\includegraphics[width=6cm]{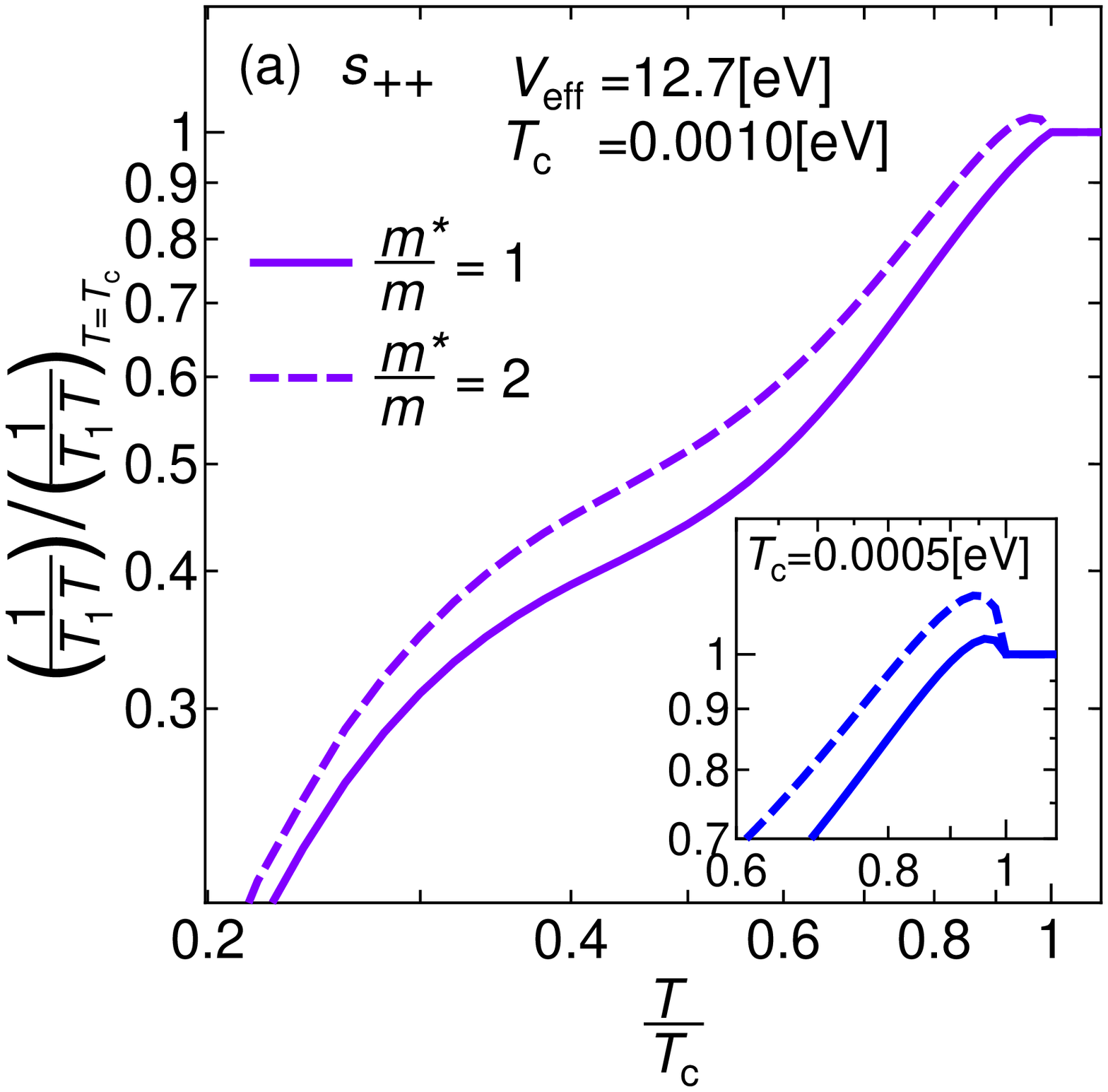}
 	\includegraphics[width=6cm]{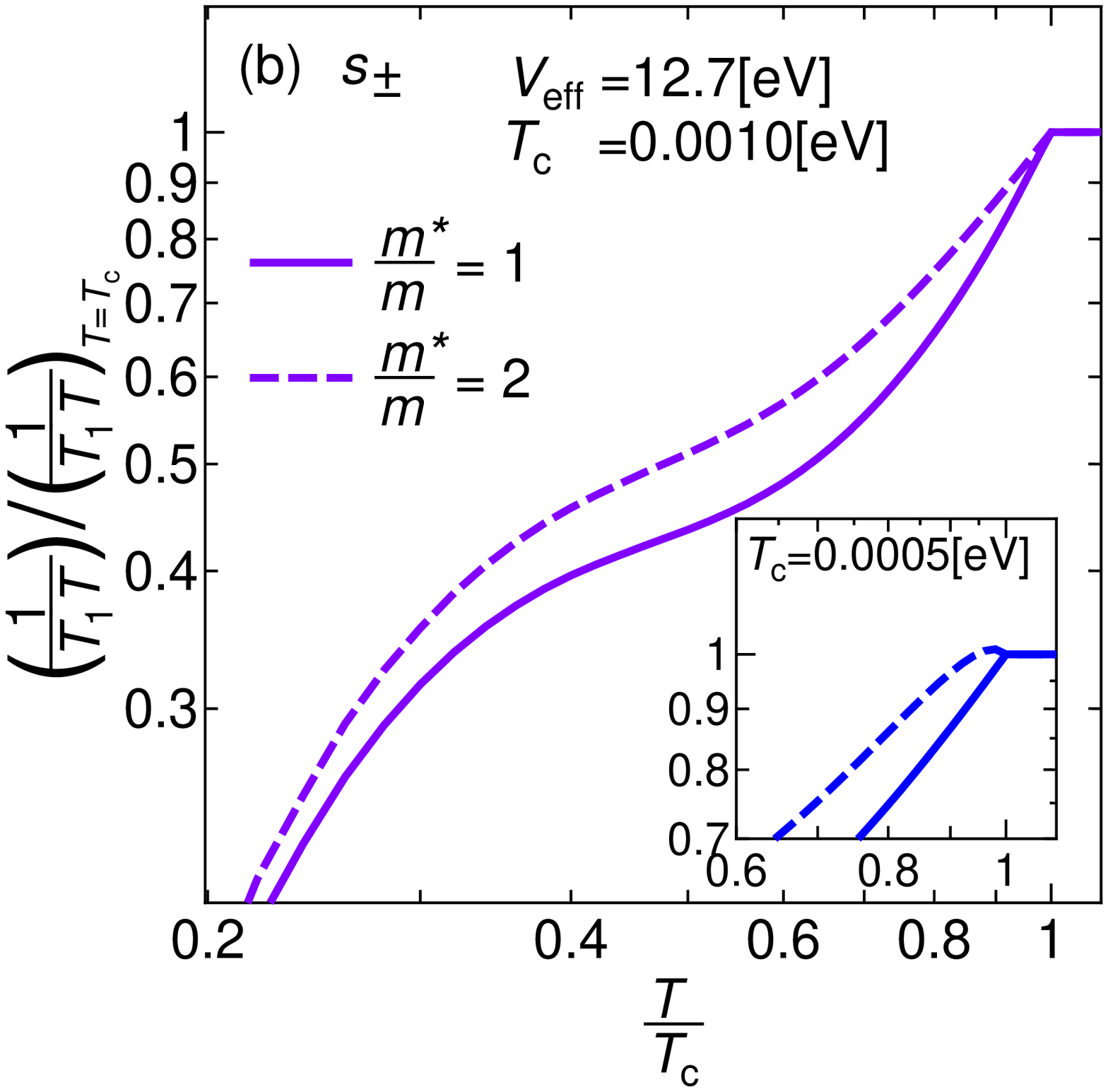}
	\caption{\label{fig:t1t_ovd} The normalized nuclear relaxation rate $(1/T_{1}T)/(1/T_{1}T)|_{T=T_{\rm c}}$ for over doping state ($T_{\rm c} = 0.0010$~eV and $V_{\rm eff} = 12.7$~eV) in the (a) $s_{++}$ and (b) $s_{\pm}$-wave states as a function of reduced temperature $T/T_{\rm c}$. 
	Solid and dashed lines represent $m^{*}/m = 1$ and $m^{*}/m = 2$, respectively. 
	$1/T_{1}T$ for $T_{\rm c} = 0.0005$~eV and $V_{\rm eff} = 12.7$~eV are shown in the insets. 
}
\end{figure}

 In Figs. \ref{fig:t1t_opt} and \ref{fig:t1t_ovd}, we show the numerical results for $T$ dependence of the nuclear spin relaxation rate for optimally and over doping states, respectively. 
 According to Table \ref{tb:gamma}, effective potential $V_{\rm eff} = 17.7$~eV for optimally doped systems and $12.7$~eV for over doped systems. 
 The mass enhancement factors are $m^{*}/m = 2 \sim 3$ for optimum doped systems and $m^{*}/m = 1 \sim 2$ for over doped system. \cite{sebastian_quantum_2008, analytis_fermi_2009, shishido_evolution_2010, qazilbash_electronic_2009, okuda_thermoelectric_2011}
 Figures \ref{fig:t1t_opt} (a) and (b) show $1/T_{1}T$ in the $s_{++}$ and $s_{\pm}$-wave states for optimally doping with $T_{\rm c} = 0.0025$~eV and $V_{\rm eff} = 17.7$~eV. 
 Solid and dashed lines represent $m^{*}/m = 2$ and $m^{*}/m = 3$, respectively. 
 In this case, the Hebel-Slichter coherence peak is suppressed even in both the $s_{\rm ++}$ and $s_{\pm}$-wave states due to the strong inelastic quasi-particle damping $\gamma^{*} (0) |_{T_{\rm c}} = 1.0 \sim 1.5 T_{\rm c}$. 

 Figures \ref{fig:t1t_ovd} (a) and (b) show $1/T_{1}T$ for over-doped systems ($T_{\rm c} = 0.0010$ eV and $V_{\rm eff} = 12.7$ eV) in $s_{++}$ and $s_{\pm}$-wave states, respectively. 
 The coherence peak increase for smaller $V_{\rm eff}$ since $\gamma$ is proportional to $V_{\rm eff}^{2}$.
 Since the coherence peak is suppressed by $\gamma^{*} = (m / m^{*}) \gamma$, it is more restored by larger $m^{*}/m$. 
 Small coherence peak is recognized in over-doped $s_{++}$-wave state for $m^{*}/m = 2$. 
 The insets of Figure \ref{fig:t1t_ovd} show $1/T_{1}T$ for $T_{\rm c} = 0.0005$~eV and $V_{\rm eff} = 12.7$~eV. 
 Since $\gamma$ decreases in proportion to $T_{\rm c}^{2}$, small coherence peak appears in both $s_{++}$ and $s_{\pm}$-wave states. 
 These results indicate that the Hebel-Slichter coherence peak on $1/T_{1}T$ might be observed for a sample with $T_{\rm c}\sim 5$~K,
as reported in Ref. \onlinecite{mukuda_coherence_2010}.

\begin{figure}[tb]
	\includegraphics[scale=0.25]{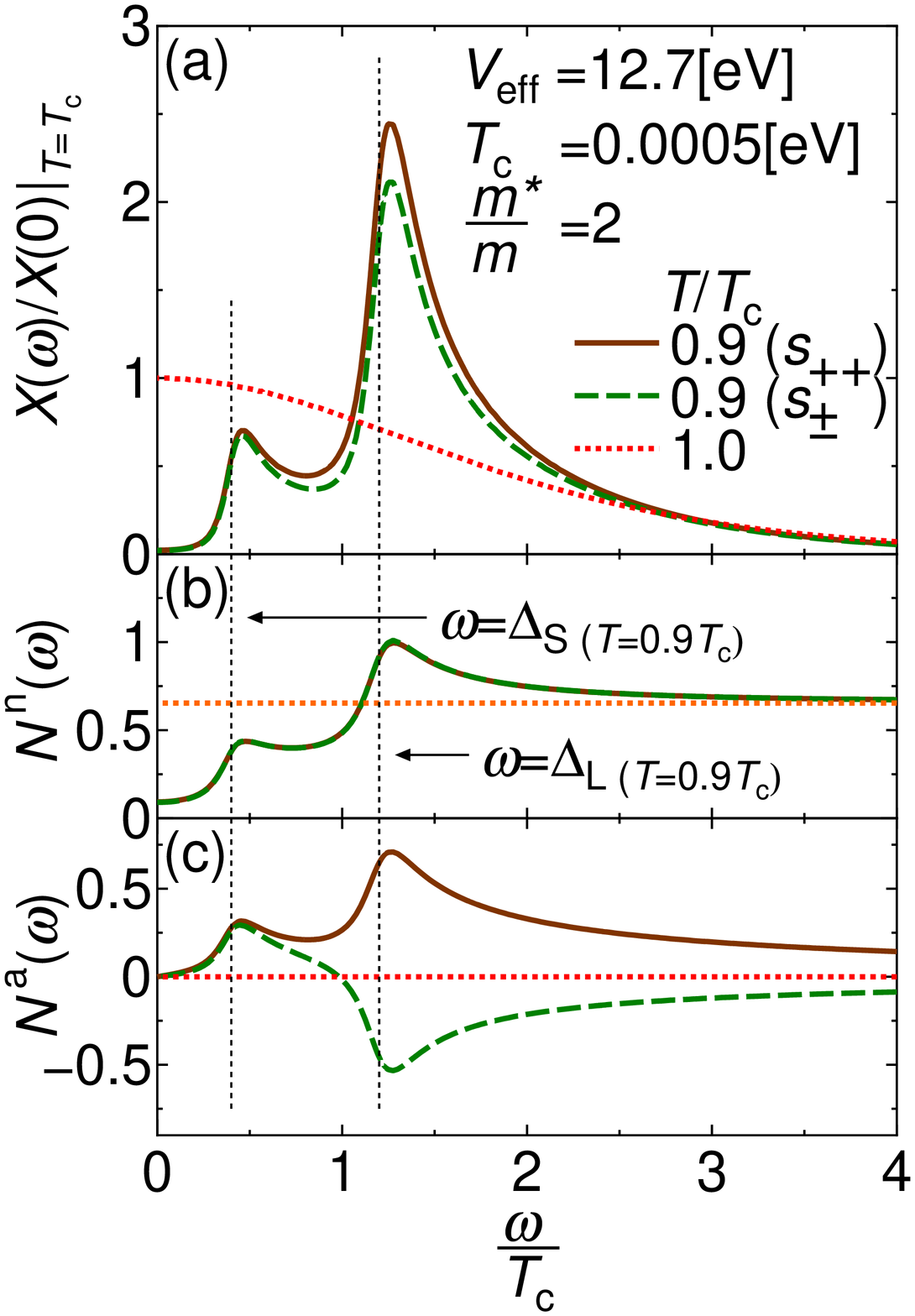}
	\includegraphics[scale=0.25]{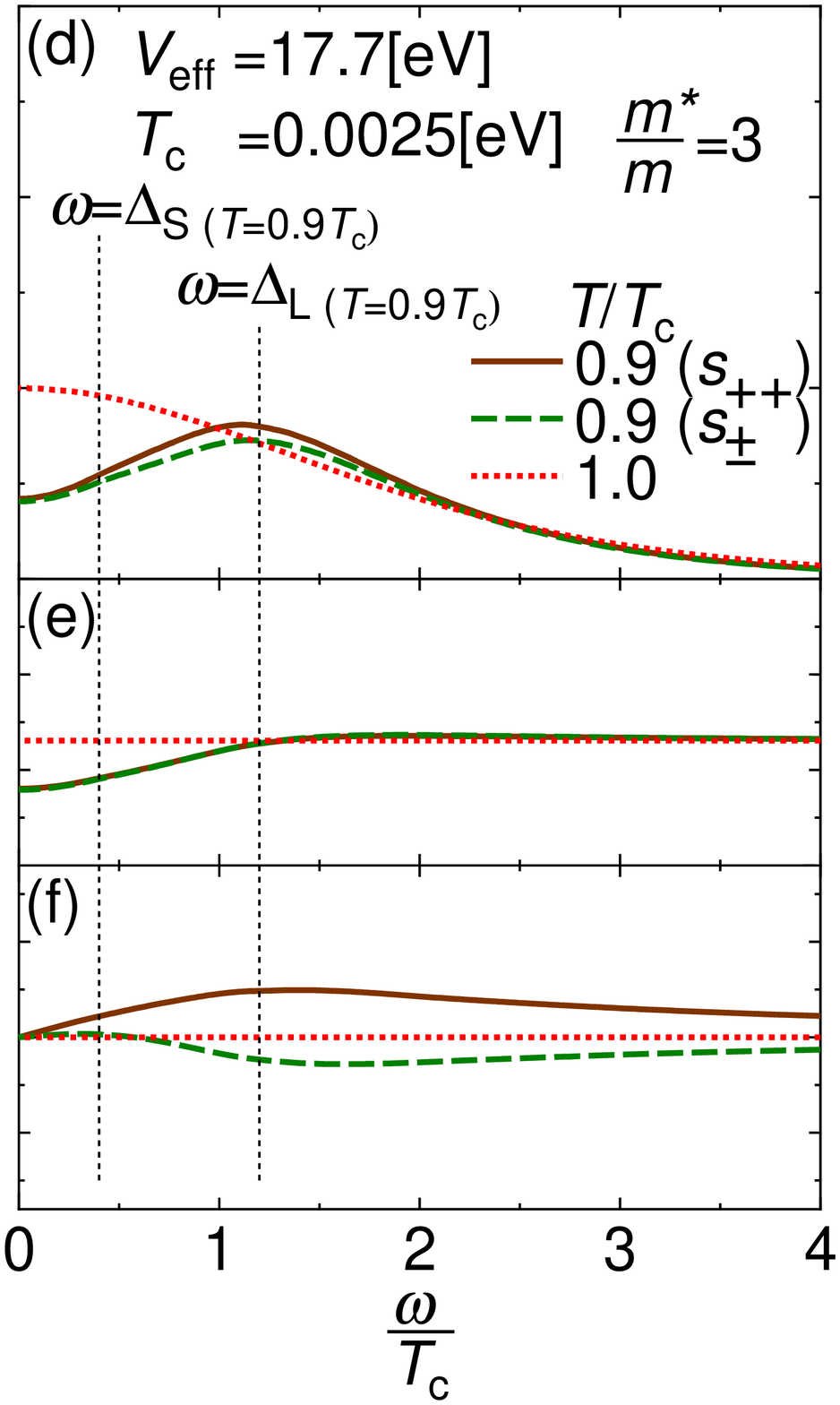}
	\caption{\label{fig:t1tw_s++} 
(a) The $\omega$ dependence of $X( \omega )/X( 0 )|_{T=T_{\rm c}}$ in over doped systems for $T_{\rm c}=0.0005$~eV, $V_{\rm eff}=12.7$~eV and $m^{*}/m = 2$. 
Solid, dashed and dotted lines represent $s_{++}$ and $s_{\pm}$-wave states ($T=0.9 T_{\rm c}$) and normal state ($T=T_{\rm c}$), respectively.
(b) $N^{\rm n}( \omega ) \equiv - {\frac{1}{\pi} \rm Im} g^{\rm R} ( \omega )$. 
(c) $N^{\rm a}( \omega ) \equiv - {\frac{1}{\pi} \rm Im} f^{\rm R} ( \omega )$. 
(d)-(f) $X ( \omega )$, $N^{\rm n}$ and $N^{\rm a}$ in optimum doped systems for $T_{\rm c} = 0.0025$, $V_{\rm eff}=17.7$~eV and $m^{*}/m = 3$. 
Vertical dotted lines show $\omega = \Delta_{\rm S}$ and $\Delta_{\rm L}$ at $T = 0.9 T_{\rm c}$. 
}
\end{figure}

 We discuss the Hebel-Slichter peak in both $s_{++}$ and $s_{\pm}$-wave states. 
 We show the $\omega$ dependences of $X ( \omega )$ (defined in Eq. (\ref{eq:t1t})), $N^{\rm n} ( \omega )$ and $N^{\rm a} ( \omega )$ for over doped systems ($T_{\rm c} = 0.0005$~eV, $V_{\rm eff} = 12.7$~eV and $m^{*}/m = 2$) in Figs. \ref{fig:t1tw_s++} (a), (b) and (c), respectively. 
 The integral of $X ( \omega ) = \pi ({N^{\rm n} ( \omega ) }^2 + {N^{\rm a} ( \omega ) }^2 ) ( - \frac{\partial f}{\partial \omega} ) $ yields $1/T_{1}T$. 
 Two peaks in $X(\omega)$ at $T = 0.9 T_{\rm c}$ correspond two superconducting gaps $\Delta_{\rm S}$ and $\pm\Delta_{\rm L}$. 
 These peaks yields the Hebel-Slichter peak in the inset of Fig. \ref{fig:t1t_ovd}. 
 For $s_{\pm}$-wave state, the sign of $N^{\rm a}$ reserves 
at $\omega \sim (\Delta_{\rm L} + \Delta_{\rm S})/2$. 
Because of the relation $|N^{\rm a}_{s_{++}}| \sim |N^{\rm a}_{s_{\pm}}|$
in the present case $\Delta_{\rm L} : \Delta_{\rm S} = 3:1$,
similar Hebel-Slichter peak appear in both superconducting states.
 On the other hand, for optimum doped systems ($T_{\rm c} = 0.0025$~eV, $V_{\rm eff} = 17.7$~eV and $m^{*}/m = 3$), these peaks in  $X ( \omega )$ are suppressed due to strong $\gamma$ at $T = 0.9 T_{\rm c}$ as shown in Figs. \ref{fig:t1tw_s++} (d) - (f).
 In this case, the Hebel-Slichter peak disappears as shown in Fig. \ref{fig:t1t_opt}. 

%%%%%%%%%%%%%%%%%%%%%%%%%%%%%%%%%%%%%%%%%%%%%%%%%%%%%%%%%%%%%%%%%%%%%%%%%%%%%%%
\section{Summary and discussion}\label{sec:summary}
 To summarize, we investigate the nuclear magnetic relaxation rate $1/T_{1}T$, especially paying attention to the presence of the coherence peak for both $s_{++}$ and $s_{\pm}$-wave states base on the five orbital model. 
 The inelastic quasi-particle damping rate $\gamma$ at $T_{\rm c}$ is estimated from the experimental results of the resistivity, and $T$- and $\omega$-dependences of $\gamma$ are calculated using the second order perturbation theory. 

 Using parameters for optimally doped systems, the Hebel-Slichter coherence peak on $1/T_{1}T$ is suppressed due to the strong  $\gamma^{*} (0) |_{T_{\rm c}} \sim T_{\rm c}$ even in the $s_{\rm ++}$-wave state for $T \ge 10$~K. 
 This result is consistent with previous strong coupling theories. \cite{akis_damping_1991, fujimoto_many-body_1992} 
 On the other hand, the relation $\gamma^{*} (0) |_{T_{\rm c}} \ll T_{\rm c}$ 
is expected in heavily over doped systems with $T_{\rm c} \sim 5$~K:
In this case, coherence peak may appear for both pairing states.
Note that the tiny coherence peak in Fig. \ref{fig:t1t_ovd} (b)
in the $s_\pm$-wave state 
grows comparable to that in \ref{fig:t1t_ovd} (a) in the $s_{++}$-wave state 
by halving the value of $\gamma^{*}(0)$.
Thus, the condition for the appearance of coherence peak 
is similar for both $s_{++}$ and $s_{\pm}$-wave states,
so it is difficult to discriminate between these pairing states
from the present NMR experimental data.

In Refs. \onlinecite{sato_studies_2010, sato_superconducting_2012},
the authors measured $1/T_{1}$ and discussed the effect of $\gamma$
on the coherence peak.
They assumed the functional form 
$\gamma ( \omega )= \gamma (0) |_{T = T_{\rm c}} \exp [A (T/T_{\rm c} - 1)]$
 ($\omega$ dependence was neglected),
 and chose parameters $A = 5$ and $\gamma (0)|_{T_{\rm c}} = 3 T_{\rm c}$,
and found that the coherence peak disappears even in the $s_{++}$-wave state.
 The large damping $\gamma(0)|_{T_{\rm c}} = 3 T_{\rm c}$ is comparable to our estimation ($1.0 \sim 1.5 T_{\rm c}$) for optimum doping systems,
so their analysis is consistent with the present study.

In this paper, we discussed only Co-doped Ba122 systems,
since reliable resistivity data in single crystals are available.
The coherence peak is also absent in LiFeAs ($c=6.36$~\AA{}, $T_{\rm c}=17$~K)
\cite{Z-Li}.
In this compound, the resistivity is fitted as
$\rho(T)=\rho_0+a T^2$ and $a=0.022$~$\mu\Omega{\rm cm/K^2}$  in Ref. \onlinecite{Heyer}.
Then, the obtained effective potential is $V_{\rm eff}=15.9$~meV,
which is comparable to the value of optimally doped Ba(Fe,Co)$_2$As$_2$ 
(17.5~meV).
Therefore, if we assume the gap structure given by ARPES measurements 
($\Delta_{\rm L}\sim \Delta_{\rm S}\sim 3$~meV) \cite{LiFeAs-ARPES1, LiFeAs-ARPES2}
the coherence peak disappears even for the $s_{++}$-wave state.
However, since the size of each hole-pocket in LiFeAs
\cite{LiFeAs-ARPES1, LiFeAs-ARPES2}
are different from that in Ba(Fe,Co)$_2$As$_2$,
more realistic tight-binding model for LiFeAs
would be required for a quantitative analysis.

In LaFeAsO$_{1-x}$F$_{x}$, 
we cannot estimate $\gamma$ quantitatively
because of the lack of resistivity data in single crystals.
For a qualitative analysis, however,
we can roughly estimated the single crystal resistivity of LaFeAsO$_{1-x}$F$_{x}$
from the poly crystal resistivity in Ref. \onlinecite{sefat_electronic_2008} 
by multiplying the factor $1/3$,
as discussed in Ref. \onlinecite{kawabata_superconductivity_2008}.
Using this method,
we can fit them by using $\rho (T)= \rho (0) + a T^{2}$ and obtain $\gamma (0) |_{T_{\rm c}} \sim 0.0023$~eV and $V_{\rm eff} \sim 10.2$~eV for LaFeAsO$_{0.89}$F$_{0.11}$ ($T_{\rm c}=26$~K). 
Using these parameters, the Hebel-Slichter coherence peak for optimally doped La1111 systems are also suppressed in both $s_{++}$ and $s_{\pm}$-wave states.

In both LiFeAs and La1111,
conventional Fermi liquid resistivity ($\rho\propto T^2$) is observed.
This fact would be consistent with 
the small spin fluctuations in these compounds. 
However, $\gamma$ at $T_{\rm c}$ is large enough to suppress the coherence peak. 
Thus, both compounds are strongly correlated Fermi liquids
due to relatively local fluctuations. 
In Sm1111 and Nd1111 ($T_{\rm c}>50$~K), 
the $T$-linear-type behavior of resistivity is observed, 
indicating the enhancement of fluctuations 
as As$_{4}$ tetrahedron approaches to a regular tetrahedron. 
\cite{saito_orbital_2010}

%add
We stress that the inelastic scattering $\gamma$
also plays important roles in the neutron scattering spectrum:
In neutron magnetic scattering measurements, 
broad "resonance-like" peak structure is observed in many 
FeAs superconductors below $T_{\rm c}$, 
and this fact had been frequently ascribed to the resonance 
due to the gap sign change. 
However, the "resonance condition" is not surely confirmed 
since it is difficult to determine the gap size accurately. 
The resonance condition in FeAs superconductor is 
$\omega_{\rm res} <\Delta_{\rm L} + \Delta_{\rm S}$, where $\omega_{\rm res}$ 
denotes the peak energy of neutron scattering.
In the following, we write down the experimental values of 
$\omega_{\rm res}$, $\Delta_{\rm L}$ and $\Delta_{\rm S}$.
\\
(A) BaFe$_{1.85}$Co$_{0.15}$As$_2$: $\omega_{\rm res}=10$~meV by neutron. \cite{Inosov}
\\
* Specific heat: \cite{hardy_doping_2010} 

$\Delta_{\rm L}= 5$~meV, $\Delta_{\rm S}=2$~meV: $\Delta_{\rm L}+\Delta_{\rm S}=7$~meV
\\
* Penetration depth: \cite{Luan} 

$\Delta_{\rm L}= 6.1$~meV, $\Delta_{\rm S}=2.3$~meV: $\Delta_{\rm L}+\Delta_{\rm S}=8.4$~meV
\\
* ARPES: \cite{Terashima} 

$\Delta_{\rm L}= 6.6$~meV, $\Delta_{\rm S}=5$~meV: $\Delta_{\rm L}+\Delta_{\rm S}=11.6$~meV
\\
\\
(B) FeTe$_{0.6}$Se$_{0.4}$: $\omega_{\rm res}=7$~meV by neutron. \cite{Harriger}
\\
* ARPES: \cite{Miao} 

$\Delta_{\rm L}=4.2$~meV, $\Delta_{\rm S}=2.5$~meV: $\Delta_{\rm L}+\Delta_{\rm S}=6.7$~meV. 
\\

In case (A), the resonance condition $\omega_{\rm res} < \Delta_{\rm L}+\Delta_{\rm S}$ 
is satisfied only by ARPES data.
In case (B), although ARPES tends to report larger gap size 
in FeAs superconductors, 
the resonance condition is still unclear.

Theoretically, we have revealed that broad hump structure at 
$\omega \sim\Delta_{\rm L} + \Delta_{\rm S}$ can appear even in the $s_{++}$-wave state, 
not due to the resonance but due to strong suppression of inelastic 
quasi-particle scattering rate $\gamma(\omega)$ for $|\omega|<3\Delta$ 
("dissipation-less mechanism" \cite{onari_structure_2010, onari_neutron_2011}).
Thus, both the absence of the coherence peak in $1/T_{1}T$ 
as well as the hump structure in the neutron scattering spectrum
are explained by the same many-body effect
--- inelastic quasi-particle scattering ---
using similar inelastic scattering rate $\gamma(0)$ 
evaluated from the resistivity.

%add
Finally, we comment that large inelastic scattering
$\gamma$ at $T_{\rm c}$ in Fe based superconductors
also works as the depairing effect.
In the random-phase-approximation (RPA),
in which the depairing effect due to inelastic scattering is neglected,
the obtained $T_{\rm c}$ is about 200~K.
\cite{kuroki_unconventional_2008,kontani_orbital-fluctuation-mediated_2010}
However, in the fluctuation-exchange (FLEX) approximation,
the obtained $T_{\rm c}$ is strongly suppressed (below 50~K)
due to the the depairing effect (self-energy correction).
\cite{onari_non-fermi-liquid_2012}

\acknowledgments
 We are grateful to M. Sato and Y. Kobayashi for useful comments and discussions. 
 This study has been supported by Grants-in-Aid for Scientific Research from MEXT of Japan, and by JST, TRIP. 
% Numerical calculations were performed at the Computer Center and the ISSP Supercomputer Center of University of Tokyo, and the Yukawa Institute Computer Facility.

%\nocite{*}

% \bibliography{nmr12}
% \bibliography{nmr8}% Produces the bibliography via BibTeX.

\begin{thebibliography}{99}
\bibitem{kamihara_iron-based_2008}
 Y. Kamihara, T. Watanabe, M. Hirano, and H. Hosono,
 J. Am. Chem. Soc. {\bf 130}, 3296 (2008).
\bibitem{hashimoto_microwave_2009}
 K. Hashimoto, T. Shibauchi, T. Kato, K. Ikada, R. Okazaki, H. Shishido, M. Ishikado, H. Kito,
 A. Iyo, H. Eisaki, S. Shamoto, and Y. Matsuda,
 Phys. Rev. Lett. {\bf 102}, 017002 (2009).
\bibitem{ding_observation_2008}
 H. Ding, P. Richard, K. Nakayama, K. Sugawara, T. Arakane, Y. Sekiba, A. Takayama, S. Souma,
 T. Sato, T. Takahashi, Z. Wang, X. Dai, Z. Fang, G. F. Chen, J. L. Luo, and N. L. Wang,
 Europhys. Lett. {\bf 83}, 47001 (2008).
\bibitem{kondo_momentum_2008}
 T. Kondo, A. F. Santander-Syro, O. Copie, C. Liu, M. E. Tillman, E. D. Mun, J. Schmalian,
 S. L. Bud'ko, M. A. Tanatar, P. C. Canfield, and A. Kaminski,
 Phys. Rev. Lett. {\bf 101}, 147003 (2008).
\bibitem{kuroki_unconventional_2008} 
 K. Kuroki, S. Onari, R. Arita, H. Usui, Y. Tanaka, H. Kontani, and H. Aoki,
 Phys. Rev. Lett. {\bf 101}, 087004 (2008).
\bibitem{mazin_unconventional_2008}
 I. I. Mazin, D. J. Singh, M. D. Johannes, and M. H. Du,
 Phys. Rev. Lett. {\bf 101}, 057003 (2008).
\bibitem{chubukov_magnetism_2008} 
 A. V. Chubukov, D. V. Efremov, and I. Eremin,
 Phys. Rev. B {\bf 78}, 134512 (2008).
\bibitem{hirschfeld_gap_2011} 
 P. J. Hirschfeld, M. M. Korshunov, and I. I. Mazin,
 Rep. Prog. Phys. {\bf 74}, 124508 (2011).
\bibitem{cvetkovic_multiband_2009} 
 V. Cvetkovic and Z. Tesanovic,
 Europhys. Lett. {\bf 85}, 37002 (2009).
\bibitem{Hanaguri}
 T. Hanaguri, S. Niitaka, K. Kuroki, and H. Takagi,
 Science {\bf 328}, 474 (2010).
\bibitem{kawabata_superconductivity_2008} 
 A. Kawabata, S. C. Lee, T. Moyoshi, Y. Kobayashi, and M. Sato,
 J. Phys. Soc. Jpn. {\bf 77}, 103704 (2008).
\bibitem{nakajima_suppression_2010} 
 Y. Nakajima, T. Taen, Y. Tsuchiya, T. Tamegai, H. Kitamura, and T. Murakami,
 Phys. Rev. B {\bf 82}, 220504 (2010).
\bibitem{li_linear_2011} 
 J. Li, Y. Guo, S. Zhang, S. Yu, Y. Tsujimoto, H. Kontani, K. Yamaura, and E. Takayama-Muromachi,
 Phys. Rev. B {\bf 84}, 020513 (2011).
\bibitem{kirshenbaum_universal_2012} 
 K. Kirshenbaum, S. R. Saha, S. Ziemak, T. Drye, and J. Paglione,
 arXiv:1203.5114.
\bibitem{onari_violation_2009} 
 S. Onari and H. Kontani,
 Phys. Rev. Lett. {\bf 103}, 177001 (2009).
\bibitem{kontani_orbital-fluctuation-mediated_2010} 
 H. Kontani and S. Onari,
 Phys. Rev. Lett. {\bf 104}, 157001 (2010).
\bibitem{onari_non-fermi-liquid_2012} 
 S. Onari and H. Kontani,
 Phys. Rev. B {\bf 85}, 134507 (2012).
\bibitem{saito_orbital_2010} 
 T. Saito, S. Onari, and H. Kontani,
 Phys. Rev. B {\bf 82}, 144510 (2010).
\bibitem{kontani_origin_2011} 
 H. Kontani, T. Saito, and S. Onari,
 Phys. Rev. B {\bf 84}, 024528 (2011).
\bibitem{onari_self-consistent_2012} 
 S. Onari and H. Kontani,
 arXiv:1203.2874.
\bibitem{kontani_orbital_2012} 
 H. Kontani, Y. Inoue, T. Saito, Y. Yamakawa, and S. Onari,
 Solid State Commun. {\bf 152}, 718 (2012).
\bibitem{fernandes_effects_2010} 
 R. M. Fernandes, L. H. VanBebber, S. Bhattacharya, P. Chandra, V. Keppens, D. Mandrus,
 M. A. McGuire, B. C. Sales, A. S. Sefat, and J. Schmalian,
 Phys. Rev. Lett. {\bf 105}, 157003 (2010).
\bibitem{goto_quadrupole_2011} 
 T. Goto, R. Kurihara, K. Araki, K. Mitsumoto, M. Akatsu, Y. Nemoto, S. Tatematsu, and M. Sato,
 J. Phys. Soc. Jpn. {\bf 80}, 073702 (2011).
\bibitem{yoshizawa_structural_2012} 
 M. Yoshizawa, D. Kimura, T. Chiba, S. Simayi, Y. Nakanishi, K. Kihou, C. Lee,A. Iyo, H. Eisaki,
 M. Nakajima, and S. Uchida,
 J. Phys. Soc. Jpn. {\bf 81}, 024604 (2012).
\bibitem{niedziela_phonon_2011}
 J. L. Niedziela, D. Parshall, K. A. Lokshin, A. S. Sefat, A. Alatas, and T. Egami,
 Phys. Rev. B {\bf 84}, 224305 (2011).
\bibitem{onari_structure_2010}
 S. Onari, H. Kontani, and M. Sato,
 Phys. Rev. B {\bf 81}, 060504 (2010).
\bibitem{onari_neutron_2011}
 S. Onari and H. Kontani,
 Phys. Rev. B {\bf 84}, 144518 (2011).
\bibitem{hebel_nuclear_1959}
 L. C. Hebel and C. P. Slichter,
 Phys. Rev. {\bf 113}, 1504 (1959).
\bibitem{nakai_evolution_2008}
 Y. Nakai, K. Ishida, Y. Kamihara, M. Hirano, and H. Hosono, 
 J. Phys. Soc. Jpn. {\bf 77}, 073701 (2008).
\bibitem{grafe_^75as_2008}
 H. Grafe,  D. Paar, G. Lang, N. J. Curro, G. Behr, J. Werner, J. Hamann-Borrero, C. Hess,
 N. Leps, R. Klingeler, and B. B\"{u}chner,
 Phys. Rev. Lett. {\bf 101}, 047003 (2008).
\bibitem{kawasaki_two_2008}
 S. Kawasaki, K. Shimada, G. F. Chen, J. L. Luo, N. L. Wang, and G.-q. Zheng,
 Phys. Rev. B {\bf 78}, 220506 (2008).
\bibitem{ning_spin_2009}
 F. Ning, K. Ahilan, T. Imai, A. S. Sefat, R. Jin, M. A. McGuire, B. C. Sales, and D. Mandrus,
 J. Phys. Soc. Jpn. {\bf 78}, 013711 (2009).
\bibitem{yashima_strong-coupling_2009}
 M. Yashima, H. Nishimura, H. Mukuda, Y. Kitaoka, K. Miyazawa, P. M. Shirage, K. Kihou, H. Kito,
 H. Eisaki, and A. Iyo,
 J. Phys. Soc. Jpn. {\bf 78}, 103702 (2009).
\bibitem{kobayashi_studies_2009}
 Y. Kobayashi, A. Kawabata, S. C. Lee, T. Moyoshi, and M. Sato,
 J. Phys. Soc. Jpn. {\bf 78}, 073704 (2009).
\bibitem{mukuda_coherence_2010}
 H. Mukuda, M. Nitta, M. Yashima, Y. Kitaoka, P. M. Shirage, H. Eisaki, and A. Iyo,
 J. Phys. Soc. Jpn. {\bf 79}, 113701 (2010).
\bibitem{akis_damping_1991}
 R. Akis and J. Carbotte,
 Solid State Commun. {\bf 78}, 393 (1991).
\bibitem{fujimoto_many-body_1992}
 S. Fujimoto,
 J. Phys. Soc. Jpn. {\bf 61}, 765 (1992).
\bibitem{kohara_superconducting_1995}
 T. Kohara, T. Oda, K. Ueda, Y. Yamada, A. Mahajan, K. Elankumaran, Z. Hossian, L. C. Gupta,
 R. Nagarajan, R. Vijayaraghavan, and C. Mazumdar,
 Phys. Rev. B {\bf 51}, 3985 (1995).
\bibitem{kishimoto_51v_1995}
 Y. Kishimoto, T. Ohno, and T. Kanashiro,
 J. Phys. Soc. Jpn. {\bf 64}, 1275 (1995).
\bibitem{bang_possible_2008}
 Y. Bang and H. Choi,
 Phys. Rev. B {\bf 78}, 134523 (2008).
\bibitem{parker_extended_2008}
 D. Parker, O. V. Dolgov, M. M. Korshunov, A. A. Golubov, and I. I. Mazin,
 Phys. Rev. B {\bf 78}, 134524 (2008).
\bibitem{senga}
 Y. Senga and H. Kontani,
 New J. Phys. {\bf 11}, 035005 (2009).
\bibitem{sato_studies_2010}
 M. Sato, Y. Kobayashi, S. C. Lee, H. Takahashi, E. Satomi, and Y. Miura,
 J. Phys. Soc. Jpn. {\bf 79}, 014710 (2010).
\bibitem{sato_superconducting_2012}
 M. Sato and Y. Kobayashi,
 Solid State Commun. {\bf 152}, 688 (2012).
\bibitem{allen_theory_1983}
 P. B. Allen and B. Mitrovi\'{c},
 {\it Theory of Superconducting Tc},
 (Solid State Physics {\bf 37}: 1-92, 1982).
\bibitem{sebastian_quantum_2008}
 S. E. Sebastian, J. Gillett, N. Harrison, P. H. C. Lau, D. J. Singh, C. H. Mielke, and
 G. G. Lonzarich,
 J. Phys.: Condens. Matter {\bf 20}, 422203 (2008).
\bibitem{analytis_fermi_2009}
 J. G. Analytis, C. M. J. Andrew, A. I. Coldea, A. McCollam, J. Chu, R. D. McDonald,
 I. R. Fisher, and A. Carrington,
 Phys. Rev. Lett. {\bf 103}, 076401 (2009).
\bibitem{shishido_evolution_2010}
 H. Shishido, A. F. Bangura, A. I. Coldea, S. Tonegawa, K. Hashimoto, S. Kasahara,
 P. M. C. Rourke, H. Ikeda, T. Terashima, R. Settai, Y. O-nuki, D. Vignolles, C. Proust,
 B. Vignolle, A. McCollam, Y. Matsuda, T. Shibauchi, and A. Carrington,
 Phys. Rev. Lett. {\bf 104}, 057008 (2010).
\bibitem{qazilbash_electronic_2009}
 M. M. Qazilbash, J. J. Hamlin, R. E. Baumbach, L. Zhang, D. J. Singh, M. B. Maple, and
 D. N. Basov,
 Nat. Phys. {\bf 5}, 647 (2009).
\bibitem{okuda_thermoelectric_2011}
 T. Okuda, W. Hirata, A. Takemori, S. Suzuki, S. Saijo, S. Miyasaka, and S. Tajima,
 J. Phys. Soc. Jpn. {\bf 80}, 044704 (2011).
\bibitem{weldon_simple_1983}
 H. A. Weldon,
 Phys. Rev. D {\bf 28}, 2007 (1983).
\bibitem{richard_fe-based_2011}
 P. Richard, T. Sato, K. Nakayama, T. Takahashi, and H. Ding,
 Rep. Prog. Phys. {\bf 74}, 124512 (2011).
\bibitem{samuely_possible_2009}
 P. Samuely, P. Szab\'o, Z. Pribulov\'a, M. E. Tillman, S. L. Bud'ko, and P. C. Canfield,
 Superc. Sci. Technol. {\bf 22}, 014003 (2009).
\bibitem{hardy_doping_2010}
 F. Hardy, P. Burger, T. Wolf, R. A. Fisher, P. Schweiss, P. Adelmann, R. Heid, R. Fromknecht,
 R. Eder, D. Ernst, H. v. L\"ohneysen, and C. Meingast,
 Europhys. Lett. {\bf 91}, 47008 (2010).
\bibitem{nagamatsu_superconductivity_2001}
 J. Nagamatsu, N. Nakagawa, T. Muranaka, Y. Zenitani, and J. Akimitsu,
 Nature {\bf 410}, 63 (2001).
\bibitem{liu_beyond_2001}
 A. Y. Liu, I. I. Mazin, and J. Kortus,
 Phys. Rev. Lett. 87, 087005 (2001).
\bibitem{johnston_puzzle_2010}
 D. C. Johnston,
 Adv. Phys. {\bf 59}, 803 (2010).
\bibitem{kontani-ROP}
 H. Kontani,
 Rep. Prog. Phys. {\bf 71}, 026501 (2008).
\bibitem{rullier-albenque_hall_2009}
 F. Rullier-Albenque, D. Colson, A. Forget, and H. Alloul,
 Phys. Rev. Lett. {\bf 103}, 057001 (2009).
\bibitem{miyake_comparison_2010}
 T. Miyake, K. Nakamura, R. Arita, and M. Imada,
 J. Phys. Soc. Jpn. {\bf 79}, 044705 (2010).
\bibitem{Sefat}
 A. S. Sefat, D. J. Singh, R. Jin, M. A. McGuire, B. C. Sales, F. Ronning, and D. Mandrus, 
 Physica C {\bf 469}, 350 (2009).
\bibitem{Z-Li}
 Z. Li, Y. Ooe, X.-C. Wang, Q.-Q. Liu, C.-Q. Jin, M. Ichioka, and G. Zheng,
 J. Phys. Soc. Jpn. {\bf 79}, 083702 (2010).
\bibitem{Heyer}
 O. Heyer, T. Lorenz, V. B. Zabolotnyy, D. V. Evtushinsky, S. V. Borisenko, I. Morozov,
 L. Harnagea, S. Wurmehl, C. Hess, and B. Buchner,
 Phys. Rev. B {\bf 84}, 064512 (2011).
\bibitem{LiFeAs-ARPES1}
 K. Umezawa, Y. Li, H. Miao, K. Nakayama, Z.-H. Liu, P. Richard, T. Sato, J. B. He,
 D.-M. Wang, G. F. Chen, H. Ding, T. Takahashi, and S.-C. Wang,
 Phys. Rev. Lett. {\bf 108}, 037002 (2012).
\bibitem{LiFeAs-ARPES2}
 S. V. Borisenko, V. B. Zabolotnyy, A. A. Kordyuk, D. V. Evtushinsky, T. K. Kim, I.
 V. Morozov, R. Follath, and B. Buchner,
 Symmetry {\bf 4}, 251 (2012).
\bibitem{sefat_electronic_2008}
 A. S. Sefat, M. A. McGuire, B. C. Sales, R. Jin, J. Y. Howe, and D. Mandrus,
 Phys. Rev. B {\bf 77}, 174503 (2008).
\bibitem{Inosov}
 D. S. Inosov, J. T. Park, P. Bourges, D. L. Sun, Y. Sidis, A. Schneidewind, K. Hradil,
 D. Haug, C. T. Lin, B. Keimer, and V. Hinkov,
 Nature Phys. {\bf 6}, 178 (2010).
\bibitem{Luan}
 L. Luan, O. M. Auslaender, T. M. Lippman, C. W. Hicks, B. Kalisky, J.-H. Chu,
 J. G. Analytis, I. R. Fisher, J. R. Kirtley, and K. A. Moler,
 Phys. Rev. B {\bf 81}, 100501 (2010).
\bibitem{Terashima}
 K. Terashima, Y. Sekiba, J. H. Bowen, K. Nakayama, T. Kawahara, T. Sato, P. Richard,
 Y.-M. Xu, L. J. Li, G. H. Cao, Z.-A. Xu, H. Ding, and T. Takahashi,
 Proc. Natl. Acad. Sci. {\bf 106}, 7330 (2009).
\bibitem{Harriger}
 L. W. Harriger, O. J. Lipscombe, C. Zhang, H. Luo, M. Wang, K. Marty, M. D. Lumsden, and P. Dai,
 Phys. Rev. B {\bf 85}, 054511 (2012).
\bibitem{Miao}
 H. Miao, P. Richard, Y. Tanaka, K. Nakayama, T. Qian, K. Umezawa, T. Sato, Y.-M. Xu,
 Y. B. Shi, N. Xu, X.-P. Wang, P. Zhang, H.-B. Yang, Z.-J. Xu, J. S. Wen, G.-D. Gu,
 X. Dai, J.-P. Hu, T. Takahashi, and H. Ding,
 Phys. Rev. B {\bf 85}, 094506 (2012).

\end{thebibliography}

\end{document}